\documentclass[prl,aps,reprint,a4paper,superscriptaddress,floatfix]{revtex4-2}

\usepackage{amsmath,amssymb,braket,dsfont,graphicx,tikz,times,xcolor}
\usetikzlibrary{patterns}

\usepackage[caption=false]{subfig}
\usepackage{hyperref}
\usepackage[capitalize,nameinlink]{cleveref}
\definecolor{darkred}{rgb}{0.5,0,0}
\definecolor{darkgreen}{rgb}{0,0.5,0}
\definecolor{darkblue}{rgb}{0,0,0.5}
\hypersetup{colorlinks=true, breaklinks=true, linkcolor=darkred, citecolor=darkgreen, urlcolor=darkblue}

\renewcommand{\arraystretch}{1.4}

\renewcommand{\leq}{\leqslant}
\renewcommand{\geq}{\geqslant}
\allowdisplaybreaks[3]

\begin{document}

\title{Exploring Bell Nonlocality with Extremal Non-Signaling Boxes}

\author{Emmanuel~Zambrini Cruzeiro}
\affiliation{\mbox{Quantum Information and Quantum Optics Laboratory, Instituto Superior Técnico, Lisboa, Portugal}}
\affiliation{Quantum Physics of Information Group, Instituto de Telecomunicações, Lisboa, Portugal}

\author{Junior~R.~Gonzales-Ureta}
\affiliation{Q*Bird BV, Delftechpark 1, 2628 XJ, Delft, The Netherlands}

\author{Raman~Choudhary}
\affiliation{INL -- International Iberian Nanotechnology Laboratory, Braga, Portugal}
\affiliation{Sorbonne Université, CNRS, LIP6, F-75005 Paris, France}

\author{Hugo~Abreu}
\affiliation{Zuse Institute Berlin, Takustraße 7, 14195 Berlin, Germany}

\author{Adán~Cabello}
\affiliation{Departamento de F\'{\i}sica Aplicada II, Universidad de Sevilla, E-41012 Sevilla, Spain}
\affiliation{Instituto Carlos~I de F\'{\i}sica Te\'orica y Computacional, Universidad de Sevilla, E-41012 Sevilla, Spain}

\author{Sébastien~Designolle}
\affiliation{Zuse Institute Berlin, Takustraße 7, 14195 Berlin, Germany}
\affiliation{Inria, ENS de Lyon, UCBL, LIP, 69342, Lyon Cedex 07, France}

\date{5th October 2026}

\begin{abstract}
  Extremal non-signaling (ENS) boxes are correlations that correspond to vertices of the non-signaling polytope of a Bell scenario.
  Neither quantum theory nor any theory of ideal measurements allows for nonlocal ENS boxes.
  That is, according to quantum theory, nonlocal ENS boxes are nonphysical.
  Still, ENS boxes are crucial for addressing a number of problems in Bell nonlocality.
  Here we obtain ENS boxes in arbitrary bipartite Bell scenarios by combining explicit constructions with computational polytope methods, and present the complete list of ENS boxes for several unexplored scenarios.
  Equipped with the boxes, we revisit several foundational questions.
  We provide a minimal decomposition of the magic square correlation---the simplest known perfect quantum correlation in nature---in terms of ENS boxes.
  For every nonlocal ENS box tested here, two copies already violate the exclusivity (or local orthogonality) principle and Specker's principle, supporting the conjecture that this is a generic feature of bipartite nonlocal ENS boxes.
  We identify the minimal scenario in which a dit of communication (with $d\leq5$) is insufficient to simulate ENS boxes.
  Our results show that the ENS-box approach leads to new results and opens new avenues for research.
\end{abstract}

\maketitle

{\em Introduction.--}
Extremal non-signaling (ENS) boxes are vertices of the non-signaling (NS) polytope $\mathcal{NS}$~\cite{Pitowsky:1989}.
Some of them are also vertices of the local polytope $\mathcal{L}$.
Neither quantum theory~\cite{Ramanathan:2010PRL} nor any theory of ideal measurements~\cite{liu2023impossibility} (which do not disturb any compatible observable and whose outcomes are confirmable) permits the existence of nonlocal vertices of the non-signaling polytope that are not local deterministic vertices $\mathcal{NS}\setminus\mathcal{L}$.
However, these boxes are of fundamental interest to physics because their convex combinations that lie outside the quantum set represent correlations that could be present in some future post-quantum theory.
Moreover, knowing the complete set of ENS boxes of a given Bell scenario is necessary to address a large number of problems.

The paradigmatic example of an ENS box is the Popescu--Rohrlich (PR) box~\cite{Popescu:1994FPH}, which is operationally defined as a black box providing 100\% success in a distribution task in a bipartite Bell scenario.
Denoting Alice's and Bob's outcomes by $a$ and $b$ and their inputs by $x$ and $y$, respectively, with all variables taking values 0 or 1, this black box perfectly satisfies
\begin{equation}
  a\oplus b=xy.
\end{equation}
In contrast, quantum strategies can win with probability at most $\frac{1}{2}(1+\frac{1}{\sqrt{2}})\approx85.36\%$.

A precise understanding of $\mathcal{NS}$ is central to the study of quantum correlations.
Several operational principles have been proposed to explain why quantum correlations are strictly weaker than NS correlations.
Thus, nonlocal ENS boxes can be used to bound the quantum set from above.
It is known that PR boxes, the only class of nonlocal ENS boxes in the simplest Bell scenario, collapse communication complexity~\cite{vanDam:1999}, violate information causality~\cite{Pawlowski:2009NAT}, macroscopic locality~\cite{Navascues:2010PRSA}, the exclusivity principle~\cite{Cabello:2013PRL}, and local orthogonality~\cite{Fritz2013}.
Nonlocal ENS boxes can shed light on the advantages and limitations of such principles.

One way of characterizing quantum correlations is to compare different resources, i.e., how they can be converted into one another.
In Bell nonlocality, one considers correlations stemming from local resources, quantum states and measurements, ENS boxes, and classical communication.
Quantum communication complexity refers to the simulation of quantum correlations using classical communication~\cite{Buhrman2010}.
Nonlocal ENS boxes can also be simulated using classical or quantum resources.
Conversely, one can simulate quantum (or any of the other) correlations using NS resources.
For example, one can simulate nonlocality arising from (partially) entangled two-qudit states~\cite{Brunner2008} using ENS boxes.
ENS boxes can be used to study how their classical and quantum~\cite{Ramanathan2022} simulation costs scale with the scenario (e.g., the minimum average communication cost), as well as the simulation of quantum behaviors using ENS boxes~\cite{Buhrman2010}.
Regarding classical communication complexity, it is known that one PR box can be simulated with one bit of communication~\cite{van2013implausible}.
This naturally raises the question: what is the smallest scenario with a nonlocal ENS box that cannot be simulated by a bit of classical communication? How about a message taking $d$ possible values, i.e., a $d$it?

\begin{figure}[ht]
  \centering
  \includegraphics[width=0.4\textwidth]{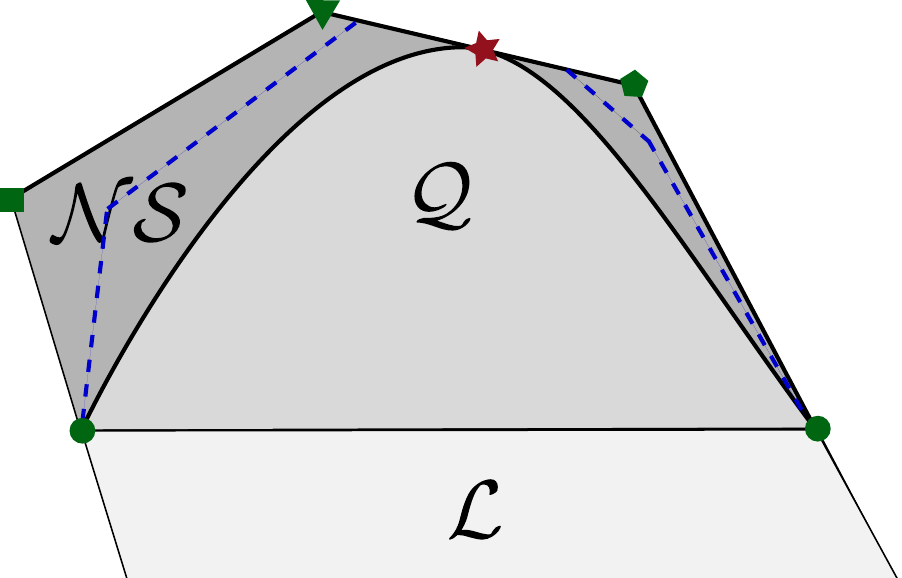}
  \caption{
  Geometric depiction of our main findings described with respect to different correlation sets: local ($\mathcal{L}$, inner set, white), quantum ($\mathcal{Q}$, light gray), and non-signaling ($\mathcal{NS}$, dark gray).
  Note that $\mathrm{LO}^{(1)}=\mathcal{NS}$, whereas $\mathrm{LO}^{(2)}$ is drawn with blue dashed lines.
  While local vertices are indicated by green disks, nonlocal non-signaling vertices are distinguished by three different green markers (square, pentagon, and triangle), illustrating the different amounts of classical communication needed to simulate these boxes.
  The red star represents magic square correlations, which we show can be decomposed into a convex combination of two ENS boxes.
  }
  \label{fig:summary}
\end{figure}

In this work, we explore Bell nonlocality using ENS boxes.
We begin by finding decompositions in terms of ENS boxes of the magic square correlations, including the minimal one consisting of just two boxes.
We then construct candidate ENS boxes by combining the methods of Jones and Masanes~\cite{JM2005PRA} and Barrett \textit{et al.}~\cite{Barrett2005}.
By checking which of the candidates are actually extremal, we can directly generate a partial list of ENS boxes in arbitrary bipartite Bell scenarios.
With this prior information, we speed up vertex enumeration, obtaining complete lists in small scenarios such as (2,3,3,2), (3,3,3,2), (2,3,3,3), and (5,4,2,2), as well as a conjecturally complete list in (3,3,3,3).
We provide a database with the lists of ENS boxes along with the article.
We find that two copies of every nonlocal ENS box tested in this work violate local orthogonality.
This is in line with what is known for previously studied bipartite ENS boxes.
Finally, we provide examples of nonlocal ENS boxes that cannot be reproduced by one $d$it of classical communication, up to $d\leq5$.
The latter results directly imply upper bounds on the number of PR boxes required to simulate a given nonlocal ENS box.
\cref{fig:summary} illustrates our results.

{\em State of the art.--}
In a bipartite Bell scenario, Alice (Bob) has $X$ ($Y$) input settings and $A$ ($B$) outcomes.
We denote such a scenario $(X,Y,A,B)$.
In the tripartite case, we add input and output cardinalities $Z$ and $C$, respectively, for the third party (Charlie).
The notation becomes $(X,Y,Z,A,B,C)$.
Complete lists of ENS boxes are available for scenarios $(d,d,2,2)$~\cite{JM2005PRA}, $(2,2,d,d)$~\cite{Barrett2005}, and $(2,2,2,2,2,2)$~\cite{Pironio2011}, for any integer $d\geq2$.
Some nonlocal ENS boxes are known in $(d,2,d,d)$~\cite{Cavalcanti2010}, and are constructed using the same ideas found in~\cite{Barrett2005}.
A general understanding of the vertices of the non-signaling polytope in arbitrary Bell scenarios is still lacking.

{\em The minimal decomposition of the magic square correlations.--}
The magic square correlations win the magic square game with probability one~\cite{Cabello:2001PRLb}.
For our purposes, it is more convenient to use the Bell inequality formulation of the magic square game~\cite{Cabello:2001PRLb}; see \cref{app:C} for details.
In this framework the magic square correlations belong to the (3,3,4,4) scenario, and they are given by

\begin{equation}
  \vec{p}_{MS}=
  \frac{1}{8}\times
  \resizebox{!}{4.5em}{$
  \left(
  \begin{array}{cccc|cccc|cccc}
    1 & 1 &   &   & 1 & 1 &   &   & 1 & 1 &   &   \\
    1 & 1 &   &   &   &   & 1 & 1 &   &   & 1 & 1 \\
      &   & 1 & 1 & 1 & 1 &   &   &   &   & 1 & 1 \\
      &   & 1 & 1 &   &   & 1 & 1 & 1 & 1 &   &   \\
    \hline
    1 &   & 1 &   & 1 &   & 1 &   &   & 1 &   & 1 \\
    1 &   & 1 &   &   & 1 &   & 1 & 1 &   & 1 &   \\
      & 1 &   & 1 & 1 &   & 1 &   & 1 &   & 1 &   \\
      & 1 &   & 1 &   & 1 &   & 1 &   & 1 &   & 1 \\
    \hline
    1 &   &   & 1 & 1 &   &   & 1 & 1 &   &   & 1 \\
    1 &   &   & 1 &   & 1 & 1 &   &   & 1 & 1 &   \\
      & 1 & 1 &   & 1 &   &   & 1 &   & 1 & 1 &   \\
      & 1 & 1 &   &   & 1 & 1 &   & 1 &   &   & 1
  \end{array}
  \right)
  $}.
\end{equation}

We begin by finding a minimal decomposition of $\vec{p}_{MS}$ in terms of ENS boxes without relying on vertex enumeration.

\textbf{Result 1.} The magic square correlations have a minimal ENS decomposition into exactly two boxes, explicitly presented below.

Suppose that the magic square correlations can be expressed as the midpoint of two extremal NS boxes, that is, there are two vectors $\vec{p}_{MS}\pm\vec{v}$ that belong to $\mathcal{NS}$ with $\vec{v}\neq\vec{0}$.
Using the positivity, normalization, and NS conditions on $\vec{p}_{MS}\pm\vec{v}$ we can derive a set of conditions for the components of $\vec{v}$.
These conditions do not completely determine $\vec{v}$, but we can choose its components to saturate as many positivity constraints as possible for $\vec{p}_{MS}\pm\vec{v}$.
With this choice of $\vec{v}$, we obtain the two correlations

\begin{equation}
  \vec{p}_1=\frac{1}{4}\times
  \resizebox{!}{4.5em}{$
  \left(
  \begin{array}{cccc|cccc|cccc}
    1 &   &   &   & 1 &   &   &   & 1 &   &   &   \\
      & 1 &   &   &   &   & 1 &   &   &   & 1 &   \\
      &   & 1 &   &   & 1 &   &   &   &   &   & 1 \\
      &   &   & 1 &   &   &   & 1 &   & 1 &   &   \\
    \hline
    1 &   &   &   & 1 &   &   &   &   & 1 &   &   \\
      &   & 1 &   &   & 1 &   &   & 1 &   &   &   \\
      & 1 &   &   &   &   & 1 &   &   &   & 1 &   \\
      &   &   & 1 &   &   &   & 1 &   &   &   & 1 \\
    \hline
    1 &   &   &   & 1 &   &   &   & 1 &   &   &   \\
      &   &   & 1 &   & 1 &   &   &   & 1 &   &   \\
      & 1 &   &   &   &   &   & 1 &   &   & 1 &   \\
      &   & 1 &   &   &   & 1 &   &   &   &   & 1
  \end{array}
  \right)
  $},\quad
  \vec{p}_2=\frac{1}{4}\times
  \resizebox{!}{4.5em}{$
  \left(
  \begin{array}{cccc|cccc|cccc}
      & 1 &   &   &   & 1 &   &   &   & 1 &   &   \\
    1 &   &   &   &   &   &   & 1 &   &   &   & 1 \\
      &   &   & 1 & 1 &   &   &   &   &   & 1 &   \\
      &   & 1 &   &   &   & 1 &   & 1 &   &   &   \\
    \hline
      &   & 1 &   &   &   & 1 &   &   &   &   & 1 \\
    1 &   &   &   &   &   &   & 1 &   &   & 1 &   \\
      &   &   & 1 & 1 &   &   &   & 1 &   &   &   \\
      & 1 &   &   &   & 1 &   &   &   & 1 &   &   \\
    \hline
      &   &   & 1 &   &   &   & 1 &   &   &   & 1 \\
    1 &   &   &   &   &   & 1 &   &   &   & 1 &   \\
      &   & 1 &   & 1 &   &   &   &   & 1 &   &   \\
      & 1 &   &   &   & 1 &   &   & 1 &   &   &
  \end{array}
  \right)
  $}
  \label{eq:compact_matrices}
\end{equation}
Moreover, it is possible to show that $\vec{p}_1$ and $\vec{p}_2$ cannot be written as a convex combination of other NS correlations, or, equivalently, that $(\vec{p}_i\pm\vec{v})\in\mathcal{NS}$ is true only when $\vec{v}=0$.
See the proof at the end of \cref{app:C}.

It is also interesting to decompose the magic square correlations in terms of PR boxes, similar to what was done in~\cite{Pironio2011}.
One possibility is to change the choice of $\vec{v}$ to aim at partial-input ENS boxes.
In this case the method does not immediately yield the decomposition, because it is not possible to decompose the magic square correlations as the midpoint of two partial-input ENS boxes.
However, after three iterations of the method, we obtain a decomposition into eight ENS vertices:

\begin{equation}
  \vec{p}_{MS}=\frac{1}{8}\sum_{i=1}^8 \vec{q}_i,\ \text{with}\quad \vec{q}_1=\frac{1}{2}\times
  \resizebox{!}{4.5em}{$
  \left(
  \begin{array}{cccc|cccc|cccc}
    2 &   &   &   & 1 & 1 &   &   & 1 & 1 &   &   \\
      &   &   &   &   &   &   &   &   &   &   &   \\
      &   &   &   &   &   &   &   &   &   &   &   \\
      &   &   &   &   &   &   &   &   &   &   &   \\
    \hline
    1 &   &   &   & 1 &   &   &   &   & 1 &   &   \\
    1 &   &   &   &   & 1 &   &   & 1 &   &   &   \\
      &   &   &   &   &   &   &   &   &   &   &   \\
      &   &   &   &   &   &   &   &   &   &   &   \\
    \hline
    1 &   &   &   & 1 &   &   &   & 1 &   &   &   \\
    1 &   &   &   &   & 1 &   &   &   & 1 &   &   \\
      &   &   &   &   &   &   &   &   &   &   &   \\
      &   &   &   &   &   &   &   &   &   &   &
  \end{array}
  \right)
  $},
\end{equation}
and the remaining $\vec{q}_i$ have a similar form.
Note that each block consists of a $4\times4$ matrix.
In this decomposition, the structure for the inputs $(x,y)=(2,2),(2,3),(3,2),(3,3)$ is a PR box for two outcomes, while the remaining entries have the structure of the boxes found in Jones and Masanes~\cite{JM2005PRA}.

{\em Vertex enumeration method and certification.--}
Several software packages specialize in facet and vertex enumeration, but PANDA offers particularly useful functionality~\cite{lorwald2015panda}.
Two features are particularly relevant here: (i) known facets or vertices can be supplied to accelerate enumeration, and (ii) PANDA directly returns symmetry classes of facets or vertices when supplied with generators of the symmetry group.
Thanks to (i) and (ii), we systematically characterize $\mathcal{NS}$ in this work, providing, in some cases, the complete list of classes of ENS boxes of a given Bell scenario.
To feed PANDA with as much data about known ENS boxes as possible, we generate candidate nonlocal ENS boxes using a structured ansatz which combines the constructions of Jones and Masanes~\cite{JM2005PRA} and Barrett \textit{et al.}~\cite{Barrett2005}.
We show below that some of these boxes are extremal; we therefore initially call them \emph{candidate} ENS boxes.
These candidates become certified ENS boxes only after their non-signaling constraints and extremality have been verified.

Formally, for $A=B$, they are defined as follows:
\begin{equation}\label{eq:1}
  \begin{pmatrix}
    S      & S         & \dots  & S           & L      & \dots  & L      \\
    S      & T_{2,2}   & \dots  & T_{2,Y-h}   & L      & \dots  & L      \\
    \vdots & \vdots    & \ddots & \vdots      & \vdots & \ddots & \vdots \\
    S      & T_{X-g,2} & \dots  & T_{X-g,Y-h} & L      & \dots  & L      \\
    K      & K         & \dots  & K           & M      & \dots  & M      \\
    \vdots & \vdots    & \ddots & \vdots      & \vdots & \ddots & \vdots \\
    K      & K         & \dots  & K           & M      & \dots  & M      \\
  \end{pmatrix},
\end{equation}
where $g\in\{0,\dots,X-2\}$ and $h\in\{0,\dots,Y-2\}$.

The matrix in \cref{eq:1} should be read as an $X\times Y$ array of $A\times A$ probability blocks.
The block in position $(x,y)$ is the matrix of conditional probabilities $p(a,b|x,y)$, whose rows are labeled by Alice's output $a$ and whose columns are labeled by Bob's output $b$.
Thus the full behavior is obtained by specifying one such $A\times A$ block for every input pair $(x,y)$.
The parameters $g$ and $h$ determine the shifts used in the block construction, while the matrices $T_{ij}$ specify the corresponding circulant/permutation structure of the $(i,j)$ block.
$T_{ij}=C_{ij}/A$ is a probability block, where $C_{ij}$ is a circulant permutation matrix.
Thus $(A\cdot T)_{ij}$ is a circulant permutation matrix.

The $A\times A$ matrix blocks are defined as follows: $S$ (resp.~$K$ and $L$) is filled with $1/A$ on the diagonal (resp.~first row and column), $M$ has a one in its first entry, and $A\cdot T_{i,j}$ is any circulant matrix, except that $A\cdot T_{2,2}$ cannot be the identity.
For $A=2$, this definition matches precisely the family defined in Ref.~\cite{JM2005PRA}, so that all these boxes are extremal.
For $A>2$, PANDA can be used to check extremality: for all the scenarios we have studied (see \cref{fig:scenarios1}), restricting to $A\cdot T_{i,j}$ being of order coprime with $A$ yields ENS boxes.
This criterion excludes $\sigma_x\otimes\mathds{1}$ for $A=4$.
We provide a small illustrative example of \cref{eq:1} in \cref{sec:exampleCandidate}.

In what follows, our workflow has four steps.
First, we generate candidate nonlocal ENS boxes using the structured block ansatz of \cref{eq:1}, inspired by Refs.~\cite{JM2005PRA,Barrett2005}.
Second, each candidate is tested for extremality in the non-signaling polytope using PANDA.
Third, the certified candidates are supplied as known vertices to PANDA, together with the symmetry generators of the Bell scenario, allowing PANDA to enumerate vertex classes.
When PANDA terminates, the obtained list is complete; otherwise, we report the corresponding list as conjectural/incomplete.
The conjectured cases occur when PANDA reaches the final stage of enumeration, where it is only checking that the already found set of inequalities is closed under adjacency.
Fourth, the resulting certified ENS boxes are used as input for the LO$^2$ and communication-complexity analyses.

{\em Complete lists of ENS boxes.--}
Using PANDA and previously known vertices, we obtain complete or partial classifications of ENS boxes in a variety of scenarios.
The scenarios are shown in \cref{fig:scenarios1,tab:scenarios2}.
We further classify the ENS boxes into full-output and partial-output vertices, in the scenarios $(2,3,3,2)$, $(3,3,3,2)$, and $(2,3,3,3)$.
This analysis is presented in \cref{app:small-scenarios}.
In some scenarios, we conjecture that the list is complete.
For example, in $(3,3,3,3)$, we find 8747 classes of boxes, which we conjecture to be the complete list of ENS boxes for that scenario.
All the results are provided in a database~\cite{GitHub}.

\textbf{Result 2.} We obtain complete and partial classifications of bipartite ENS boxes in several Bell scenarios, including examples beyond PR-box-based constructions.

{\em Violations of local orthogonality.--}
Bounding the quantum set of nonlocal correlations from information-theoretic and physical principles is an open problem~\cite{Cabello:2001PRLb,Cabello:2013PRL}.
One principle that is useful for this task is local orthogonality (LO), which is the name given in the literature to the exclusivity principle~\cite{Cabello:2013PRL,acin2015combinatorial} in Bell scenarios.
In some frameworks, the exclusivity principle is not a principle but a theorem.
For example, it has to be satisfied by theories of ideal measurements~\cite{cabello2019quantum,chiribella2020general}.
Specker's principle implies the exclusivity principle~\cite{cabello2163contextuality}, which is called LO in Bell scenarios~\cite{Fritz2013,sainz2014exploring}.
In~\cite{sainz2014exploring,Fritz2013}, the authors explored all nonlocal ENS boxes of bipartite scenarios (2,2,2,2), ($m$,$m$,2,2), (2,2,$d$,$d$), and the tripartite scenario (2,2,2,2,2,2) to obtain violations of LO.
Their main result can be summarized as follows: each of the bipartite ENS boxes that they considered necessarily requires two statistically independent copies of the box to violate the LO principle, while only one copy of the tripartite box violates an LO inequality.
The former is expressed as the violation of the LO$^2$ inequality while the latter is defined as the violation of an LO$^1$ inequality.
A natural question is: can we find examples of nonlocal ENS correlations which satisfy every LO$^2$ inequality?
It has already been shown in~\cite{sainz2014exploring,cabello2163contextuality} that for any single copy of a bipartite scenario the set of correlations satisfying LO$^1$ is the same as the set of correlations satisfying the NS principle.
Therefore, for each nonlocal ENS box found in this work, we look for a violation of some LO$^2$ inequality.

Finding a violation of an LO$^k$ inequality amounts to finding a clique (complete graph) whose total probability weight exceeds one in the corresponding exclusivity graph.
This exclusivity graph, called the joint exclusivity graph, is obtained by taking the OR product of $k$ copies of the exclusivity graph of the correlation under consideration.
For a given correlation, an exclusivity graph is defined by vertices that represent possible events ($ab|XY$) in the correlation, node weights that represent probabilities of events, i.e., $P(ab|XY)$, and edges that mark the connected events as exclusive: two nodes $ab|XY$ and $a'b'|X'Y'$ are exclusive if and only if, whenever $X=X'$, then $a\neq a'$, or whenever $Y=Y'$, then $b\neq b'$.
Taking the OR product of $k$ such graphs gives us the joint graph where each node denotes a joint event represented by a tuple consisting of $k$ events, one from each copy, each node weight is the product of probabilities of the corresponding $k$ events, and each edge between two nodes denotes exclusivity defined by the exclusivity of at least one pair of tuple components across the two nodes.
In other words, the exclusivity across two nodes in the joint exclusivity graph is due to two events coming from (at least) one of the individual exclusivity graphs.
Note that it is the clique size as well as node weights that determine the violation.
For extremal boxes such as the PR box, where all possible events occur with the same probability, i.e., the corresponding graph has uniform node weights, the violation only depends on the size of the clique but, more generally, both the clique size and its node weights determine violations of the corresponding LO inequalities.

For example, consider the ENS box given in \cref{eq:box-1}.
Focusing on LO$^2$ means considering two statistically independent copies of this box.
This means that if $e_{i}^1$ and $e_{j}^2$ are possible events from two different copies of the extremal box, and the joint event is denoted by $(e_i^{1},e_j^{2})$, then $P(e_i^{1},e_j^{2})=P(e_i^{1}) P(e_j^{2})$.

For the extremal box in \cref{eq:box-1}, there are only three nontrivial possible values of the joint probability of any event $(e_i^{1},e_j^{2})$: $\frac{1}{4},\frac{1}{16}$, and $\frac{1}{8}$.
$\frac{1}{4}$ occurs when $P(e_i^{1})=P(e_j^{2})=\frac{1}{2}$, $\frac{1}{16}$ when $P(e_i^{1})=P(e_j^{2})=\frac{1}{4}$, and $\frac{1}{8}$ when either $P(e_i^{1})$ or $P(e_j^{2})$ is $\frac{1}{4}$ while the other one is $\frac{1}{2}$.
Consider a clique whose vertices are joint events, each comprising one event from each of the two independent copies.
If it has $x$ vertices of probability $1/4$, $y$ of probability $1/16$, and $z$ of probability $1/8$, then its LO$^2$ inequality is violated precisely when
\begin{equation}
  \frac{x}{4}+\frac{y}{16}+\frac{z}{8}>1,
  \label{eq:36}
\end{equation}
or equivalently, $4x+y+2z>16$.
In general, if $k$ is the size of the clique, i.e., $x+y+z=k$, then any solution to this equation must satisfy the following condition to violate the LO$^2$ inequality: $3x+z>16-k$.
It can be algebraically argued that no solution to the inequality $x+y+z\leq4$ violates inequality \eqref{eq:36}.
Therefore, the smallest clique potentially violating the LO$^2$ inequality has size $k\geq5$.
It can be further logically argued that no clique of size $k=5$ would yield a violation either.
Informed by this, we performed a computational search for a clique violating the LO$^2$ inequality and identified a $K_9$ that yields such a violation.
The set of these nine locally orthogonal events is $\{1120|0011,\allowbreak 1001|1212,\allowbreak 2010|0010,\allowbreak 0121|1021,\allowbreak 2101|0220,\allowbreak 2020|0010,\allowbreak 2111|0220,\allowbreak 0101|0221,\allowbreak 0120|0011\}$.
For this $K_9$, the counts in \cref{eq:36} are $x=1$, $y=3$, and $z=5$, giving $1/4+3/16+5/8=17/16>1$.

The same analysis is provided for other examples of boxes in \cref{app:LO2}.
For the boxes considered with binary outcomes, the violated inequality LO$^2$ always corresponds to a $K_5$.
The same is true for binary-outcome bipartite ENS boxes considered in~\cite{Fritz2013}.
Furthermore, for all tested classes of nonlocal ENS boxes with more than binary outcomes, we also find a violation of some LO$^2$ inequality.

\textbf{Result 3.} Every nonlocal ENS box found in this work violates LO at the two-copy level, LO$^2$.

We conjecture that in bipartite Bell scenarios, two copies of any nonlocal ENS box always violate some LO$^2$ inequality.

\textbf{Conjecture 1.} Every bipartite nonlocal ENS box violates LO$^2$.

Note that this is not true in general: the tripartite ENS box for (2,2,2,2,2,2) mentioned previously does violate an LO$^1$ inequality, and consequently an LO$^2$ inequality as well, whereas extremal boxes for certain Kochen--Specker scenarios do not even violate inequalities corresponding to LO$^3$ as shown in~\cite{choudhary2024exclusivity}.

{\em Communication complexity of ENS boxes.--}
A fundamental step for understanding why some correlations are impossible is quantifying their simulation cost.
Here, we present new results on the amount of classical communication required to simulate ENS boxes.
Using the Frank--Wolfe technique in~\cite{designolle2023improved,braun2022conditional} adapted for local hidden variable models supplemented with classical communication, we systematically look for violations of the one-$d$it polytope.
In our notation, a $d$it means a single classical message $m\in\{1,\ldots,d\}$, i.e., one use of a noiseless $d$-ary classical channel.
The communication is one-way, from Alice to Bob, after Alice receives her input and before Bob produces his output.
Alice and Bob may share arbitrary prior randomness.
The resulting set, denoted LHV+$d$it, is the convex hull of deterministic behaviors
\begin{equation}
  p(a,b|x,y)=\sum_\lambda q_\lambda\,\delta_{a,f_\lambda(x)}\delta_{b,g_\lambda(y,m_\lambda(x))},
\end{equation}
where $m_\lambda(x)\in\{1,\ldots,d\}$, and $f_\lambda,g_\lambda$ are deterministic maps for Alice and Bob, respectively.
The communication cost is worst-case: for every run and every input $x$, Alice sends exactly one $d$-valued message, or equivalently at most one such message.
In \cref{tab:comm}, we provide the number of boxes which violate each case, along with the minimum visibility for that scenario.
This allows us to provide minimal examples of ENS boxes that cannot be simulated by a $d$it of communication, for $d\leq5$.

\textbf{Result 4.} The ENS boxes studied here admit finite one-way classical-communication simulations, with message sizes ranging from one classical bit to a five-level classical message, i.e., a quint.

The highest communication complexity is achieved in the scenario $(6,4,2,2)$.
Along with these behaviors, we provide facets separating them from the polytope of local correlations with supplementary classical $d$it communication.

We find that the (3,3,3,3) ENS box first presented in~\cite{JM2005PRA} can be simulated with one bit of communication.
In contrast, 8160 out of the 8747 classes of boxes cannot be reproduced with one bit, i.e., require a trit.

It turns out that the scenario $(m,m,2,2)$, already solved in~\cite{JM2005PRA}, allows us to generalize the pattern that emerges from \cref{tab:comm22}.
In this scenario, consider the following extremal NS probability distribution (in the matrix-matrix notation from~\cite{JM2005PRA})
\begin{equation}
  p=\left(\begin{array}{c|c|c|c}
    C      & C      & \cdots & C      \\ \hline
    C      & A      & \cdots & C      \\ \hline
    \vdots & \vdots & \ddots & \vdots \\ \hline
    C      & C      & \cdots & A      \\
  \end{array}\right),
\end{equation}
where $C=\frac12\bigl(\begin{smallmatrix}1 & 0\\0 & 1\end{smallmatrix}\bigr)$ and $A=\frac12\bigl(\begin{smallmatrix}0 & 1\\1 & 0\end{smallmatrix}\bigr)$.
Define $F=4p-1$; the value attained by $p$ for this inequality is $m^2$, while we prove in \cref{app:dit} that the bounds with communication $d\leq m$ are $L_d=m^2-2(m-d)$.
This shows that $p$ is not in LHV+$(m-1)$it.
Note that the inequality $F$ generalizes CHSH in a way that strongly favors NS strategies at the expense of quantum ones.
The quantum value is indeed $2\sqrt2$ for $m=2$, 6 for $m=3$ (still nonlocal), but it then coincides with the local bound for $m\geq4$.
A proof of these quantum bounds is given in \cref{app:dit}.

{\em Discussion.--}
PR boxes have been a crucial tool for understanding quantum correlations and framing more general theories.
However, PR boxes are just \emph{one} example of ENS boxes.
Each Bell scenario has its own set.
Therefore, the same reasons why PR boxes are of interest lead to the fundamental question of what the corresponding tools in more general scenarios are, why these tools do not exist, what their simulation cost is, and how existing correlations can be understood in terms of ENS boxes.
In this Letter, we have introduced methods and results in all these areas that we expect to open a fruitful path for research and provide a new perspective on quantum correlations.

We show that ENS boxes beyond the PR-box paradigm provide a systematic resource-theoretic probe of Bell nonlocality: they yield new constraints from exclusivity/LO, reveal minimal classical-communication simulation costs, and provide minimal decompositions of canonical perfect quantum correlations such as the magic square.
The central message is not the enumeration itself, but that access to ENS vertices beyond the PR-box paradigm converts several qualitative questions about Bell nonlocality into concrete, solvable resource-theoretic problems.

The results presented here offer several promising directions to address other fascinating open problems.
First, they can be used to generate new Bell inequalities using the techniques in~\cite{Brunner2006,Wolfe2012,Cope2019} and to find minimal ENS box decompositions of fully nonlocal, post-quantum correlations~\cite{Ravi2021PRR}.
Second, we have only explored one principle---the exclusivity principle/LO---for bounding the quantum set from above using NS correlations.
Several other principles offer rich avenues for further investigation.
Finally, ENS boxes also have important applications in quantum communication, since they can be used to generate randomness from arbitrarily weak seeds, which requires Hardy-type behavior or pseudotelepathy~\cite{Liu2023}, and are valuable in QKD security proofs under the worst-case assumption of a no-signaling adversary~\cite{Masanes2014}.

{\em Acknowledgments.--}
E.Z.C.~acknowledges funding from FCT/MCTES---Fundação para a Ciência e a Tecnologia (Portugal)---through national funds and, when applicable, cofunding from EU funds under project UIDB/50008/2020.
A.C.~was supported by the EU-funded project FoQaCiA.
H.A.~and S.D.~were supported through the Research Campus Modal funded by the German Federal Ministry of Education and Research (fund numbers 05M14ZAM, 05M20ZBM).
S.D.~was partly funded by the QuantERA II Programme that has received funding from the European Union's Horizon 2020 research and innovation program under Grant Agreement No.~101017733 (VERIqTAS).
R.C.~acknowledges support from FCT (SFRH/BD/151452/2021) and ANR project QNS (ANR-24-CE97-0005-01).
GPT 6 Astra was used to correct minor grammatical and mathematical errors.

\bibliographystyle{apsrev4-2}
\bibliography{ZGC+26}

\begin{thebibliography}{34}%
\makeatletter
\providecommand \@ifxundefined [1]{%
 \@ifx{#1\undefined}
}%
\providecommand \@ifnum [1]{%
 \ifnum #1\expandafter \@firstoftwo
 \else \expandafter \@secondoftwo
 \fi
}%
\providecommand \@ifx [1]{%
 \ifx #1\expandafter \@firstoftwo
 \else \expandafter \@secondoftwo
 \fi
}%
\providecommand \natexlab [1]{#1}%
\providecommand \enquote  [1]{``#1''}%
\providecommand \bibnamefont  [1]{#1}%
\providecommand \bibfnamefont [1]{#1}%
\providecommand \citenamefont [1]{#1}%
\providecommand \href@noop [0]{\@secondoftwo}%
\providecommand \href [0]{\begingroup \@sanitize@url \@href}%
\providecommand \@href[1]{\@@startlink{#1}\@@href}%
\providecommand \@@href[1]{\endgroup#1\@@endlink}%
\providecommand \@sanitize@url [0]{\catcode `\\12\catcode `\$12\catcode `\&12\catcode `\#12\catcode `\^12\catcode `\_12\catcode `\%12\relax}%
\providecommand \@@startlink[1]{}%
\providecommand \@@endlink[0]{}%
\providecommand \url  [0]{\begingroup\@sanitize@url \@url }%
\providecommand \@url [1]{\endgroup\@href {#1}{\urlprefix }}%
\providecommand \urlprefix  [0]{URL }%
\providecommand \Eprint [0]{\href }%
\providecommand \doibase [0]{https://doi.org/}%
\providecommand \selectlanguage [0]{\@gobble}%
\providecommand \bibinfo  [0]{\@secondoftwo}%
\providecommand \bibfield  [0]{\@secondoftwo}%
\providecommand \translation [1]{[#1]}%
\providecommand \BibitemOpen [0]{}%
\providecommand \bibitemStop [0]{}%
\providecommand \bibitemNoStop [0]{.\EOS\space}%
\providecommand \EOS [0]{\spacefactor3000\relax}%
\providecommand \BibitemShut  [1]{\csname bibitem#1\endcsname}%
\let\auto@bib@innerbib\@empty
\bibitem [{\citenamefont {Pitowsky}(1989)}]{Pitowsky:1989}%
  \BibitemOpen
  \bibfield  {author} {\bibinfo {author} {\bibfnamefont {I.}~\bibnamefont {Pitowsky}},\ }\href {https://doi.org/10.1007/BFb0021186} {\emph {\bibinfo {title} {Quantum Probability---Quantum Logic}}},\ \bibinfo {series} {Lecture Notes in Physics}, Vol.\ \bibinfo {volume} {321}\ (\bibinfo  {publisher} {Springer-Verlag},\ \bibinfo {address} {Berlin},\ \bibinfo {year} {1989})\BibitemShut {NoStop}%
\bibitem [{\citenamefont {Ramanathan}\ \emph {et~al.}(2016)\citenamefont {Ramanathan}, \citenamefont {Tuziemski}, \citenamefont {Horodecki},\ and\ \citenamefont {Horodecki}}]{Ramanathan:2010PRL}%
  \BibitemOpen
  \bibfield  {author} {\bibinfo {author} {\bibfnamefont {R.}~\bibnamefont {Ramanathan}}, \bibinfo {author} {\bibfnamefont {J.}~\bibnamefont {Tuziemski}}, \bibinfo {author} {\bibfnamefont {M.}~\bibnamefont {Horodecki}},\ and\ \bibinfo {author} {\bibfnamefont {P.}~\bibnamefont {Horodecki}},\ }\href {https://doi.org/10.1103/PhysRevLett.117.050401} {\bibfield  {journal} {\bibinfo  {journal} {Phys. Rev. Lett.}\ }\textbf {\bibinfo {volume} {117}},\ \bibinfo {pages} {050401} (\bibinfo {year} {2016})}\BibitemShut {NoStop}%
\bibitem [{\citenamefont {Liu}\ \emph {et~al.}(2024{\natexlab{a}})\citenamefont {Liu}, \citenamefont {Chung}, \citenamefont {Cruzeiro}, \citenamefont {Gonzales-Ureta}, \citenamefont {Ramanathan},\ and\ \citenamefont {Cabello}}]{liu2023impossibility}%
  \BibitemOpen
  \bibfield  {author} {\bibinfo {author} {\bibfnamefont {Y.}~\bibnamefont {Liu}}, \bibinfo {author} {\bibfnamefont {H.~Y.}\ \bibnamefont {Chung}}, \bibinfo {author} {\bibfnamefont {E.~Z.}\ \bibnamefont {Cruzeiro}}, \bibinfo {author} {\bibfnamefont {J.~R.}\ \bibnamefont {Gonzales-Ureta}}, \bibinfo {author} {\bibfnamefont {R.}~\bibnamefont {Ramanathan}},\ and\ \bibinfo {author} {\bibfnamefont {A.}~\bibnamefont {Cabello}},\ }\href {https://doi.org/10.1103/PhysRevResearch.6.L042035} {\bibfield  {journal} {\bibinfo  {journal} {Phys. Rev. Res.}\ }\textbf {\bibinfo {volume} {6}},\ \bibinfo {pages} {L042035} (\bibinfo {year} {2024}{\natexlab{a}})}\BibitemShut {NoStop}%
\bibitem [{\citenamefont {Popescu}\ and\ \citenamefont {Rohrlich}(1994)}]{Popescu:1994FPH}%
  \BibitemOpen
  \bibfield  {author} {\bibinfo {author} {\bibfnamefont {S.}~\bibnamefont {Popescu}}\ and\ \bibinfo {author} {\bibfnamefont {D.}~\bibnamefont {Rohrlich}},\ }\href {https://doi.org/10.1007/BF02058098} {\bibfield  {journal} {\bibinfo  {journal} {Found. Phys.}\ }\textbf {\bibinfo {volume} {24}},\ \bibinfo {pages} {379} (\bibinfo {year} {1994})}\BibitemShut {NoStop}%
\bibitem [{\citenamefont {van Dam}(1999)}]{vanDam:1999}%
  \BibitemOpen
  \bibfield  {author} {\bibinfo {author} {\bibfnamefont {W.}~\bibnamefont {van Dam}},\ }\emph {\bibinfo {title} {Nonlocality \& Communication Complexity}},\ \href {https://sites.cs.ucsb.edu/~vandam/oxford_thesis.pdf} {Ph.D. thesis},\ \bibinfo  {school} {University of Oxford} (\bibinfo {year} {1999}),\ \bibinfo {note} {chapter 9}\BibitemShut {NoStop}%
\bibitem [{\citenamefont {Paw{\l}owski}\ \emph {et~al.}(2009)\citenamefont {Paw{\l}owski}, \citenamefont {Paterek}, \citenamefont {Kaszlikowski}, \citenamefont {Scarani}, \citenamefont {Winter},\ and\ \citenamefont {{\.Z}ukowski}}]{Pawlowski:2009NAT}%
  \BibitemOpen
  \bibfield  {author} {\bibinfo {author} {\bibfnamefont {M.}~\bibnamefont {Paw{\l}owski}}, \bibinfo {author} {\bibfnamefont {T.}~\bibnamefont {Paterek}}, \bibinfo {author} {\bibfnamefont {D.}~\bibnamefont {Kaszlikowski}}, \bibinfo {author} {\bibfnamefont {V.}~\bibnamefont {Scarani}}, \bibinfo {author} {\bibfnamefont {A.}~\bibnamefont {Winter}},\ and\ \bibinfo {author} {\bibfnamefont {M.}~\bibnamefont {{\.Z}ukowski}},\ }\href {https://doi.org/10.1038/nature08400} {\bibfield  {journal} {\bibinfo  {journal} {Nature}\ }\textbf {\bibinfo {volume} {461}},\ \bibinfo {pages} {1101} (\bibinfo {year} {2009})}\BibitemShut {NoStop}%
\bibitem [{\citenamefont {Navascu{\'e}s}\ and\ \citenamefont {Wunderlich}(2010)}]{Navascues:2010PRSA}%
  \BibitemOpen
  \bibfield  {author} {\bibinfo {author} {\bibfnamefont {M.}~\bibnamefont {Navascu{\'e}s}}\ and\ \bibinfo {author} {\bibfnamefont {H.}~\bibnamefont {Wunderlich}},\ }\href {https://doi.org/10.1098/rspa.2009.0453} {\bibfield  {journal} {\bibinfo  {journal} {Proc. R. Soc. A}\ }\textbf {\bibinfo {volume} {466}},\ \bibinfo {pages} {881} (\bibinfo {year} {2010})}\BibitemShut {NoStop}%
\bibitem [{\citenamefont {Cabello}(2013)}]{Cabello:2013PRL}%
  \BibitemOpen
  \bibfield  {author} {\bibinfo {author} {\bibfnamefont {A.}~\bibnamefont {Cabello}},\ }\href {https://doi.org/10.1103/PhysRevLett.110.060402} {\bibfield  {journal} {\bibinfo  {journal} {Phys. Rev. Lett.}\ }\textbf {\bibinfo {volume} {110}},\ \bibinfo {pages} {060402} (\bibinfo {year} {2013})}\BibitemShut {NoStop}%
\bibitem [{\citenamefont {Fritz}\ \emph {et~al.}(2013)\citenamefont {Fritz}, \citenamefont {Sainz}, \citenamefont {Augusiak}, \citenamefont {Brask}, \citenamefont {Chaves}, \citenamefont {Leverrier},\ and\ \citenamefont {Ac{\'\i}n}}]{Fritz2013}%
  \BibitemOpen
  \bibfield  {author} {\bibinfo {author} {\bibfnamefont {T.}~\bibnamefont {Fritz}}, \bibinfo {author} {\bibfnamefont {A.~B.}\ \bibnamefont {Sainz}}, \bibinfo {author} {\bibfnamefont {R.}~\bibnamefont {Augusiak}}, \bibinfo {author} {\bibfnamefont {J.~B.}\ \bibnamefont {Brask}}, \bibinfo {author} {\bibfnamefont {R.}~\bibnamefont {Chaves}}, \bibinfo {author} {\bibfnamefont {A.}~\bibnamefont {Leverrier}},\ and\ \bibinfo {author} {\bibfnamefont {A.}~\bibnamefont {Ac{\'\i}n}},\ }\href {https://doi.org/10.1038/ncomms3263} {\bibfield  {journal} {\bibinfo  {journal} {Nat. Commun.}\ }\textbf {\bibinfo {volume} {4}},\ \bibinfo {pages} {2263} (\bibinfo {year} {2013})}\BibitemShut {NoStop}%
\bibitem [{\citenamefont {Buhrman}\ \emph {et~al.}(2010)\citenamefont {Buhrman}, \citenamefont {Cleve}, \citenamefont {Massar},\ and\ \citenamefont {de~Wolf}}]{Buhrman2010}%
  \BibitemOpen
  \bibfield  {author} {\bibinfo {author} {\bibfnamefont {H.}~\bibnamefont {Buhrman}}, \bibinfo {author} {\bibfnamefont {R.}~\bibnamefont {Cleve}}, \bibinfo {author} {\bibfnamefont {S.}~\bibnamefont {Massar}},\ and\ \bibinfo {author} {\bibfnamefont {R.}~\bibnamefont {de~Wolf}},\ }\href {https://doi.org/10.1103/RevModPhys.82.665} {\bibfield  {journal} {\bibinfo  {journal} {Rev. Mod. Phys.}\ }\textbf {\bibinfo {volume} {82}},\ \bibinfo {pages} {665} (\bibinfo {year} {2010})}\BibitemShut {NoStop}%
\bibitem [{\citenamefont {Brunner}\ \emph {et~al.}(2008)\citenamefont {Brunner}, \citenamefont {Gisin}, \citenamefont {Popescu},\ and\ \citenamefont {Scarani}}]{Brunner2008}%
  \BibitemOpen
  \bibfield  {author} {\bibinfo {author} {\bibfnamefont {N.}~\bibnamefont {Brunner}}, \bibinfo {author} {\bibfnamefont {N.}~\bibnamefont {Gisin}}, \bibinfo {author} {\bibfnamefont {S.}~\bibnamefont {Popescu}},\ and\ \bibinfo {author} {\bibfnamefont {V.}~\bibnamefont {Scarani}},\ }\href {https://doi.org/10.1103/PhysRevA.78.052111} {\bibfield  {journal} {\bibinfo  {journal} {Phys. Rev. A}\ }\textbf {\bibinfo {volume} {78}},\ \bibinfo {pages} {052111} (\bibinfo {year} {2008})}\BibitemShut {NoStop}%
\bibitem [{\citenamefont {Ramanathan}\ \emph {et~al.}(2022)\citenamefont {Ramanathan}, \citenamefont {Banacki}, \citenamefont {Ravell~Rodr{\'\i}guez},\ and\ \citenamefont {Horodecki}}]{Ramanathan2022}%
  \BibitemOpen
  \bibfield  {author} {\bibinfo {author} {\bibfnamefont {R.}~\bibnamefont {Ramanathan}}, \bibinfo {author} {\bibfnamefont {M.}~\bibnamefont {Banacki}}, \bibinfo {author} {\bibfnamefont {R.}~\bibnamefont {Ravell~Rodr{\'\i}guez}},\ and\ \bibinfo {author} {\bibfnamefont {P.}~\bibnamefont {Horodecki}},\ }\href {https://doi.org/10.1038/s41534-022-00633-0} {\bibfield  {journal} {\bibinfo  {journal} {npj Quantum Inf.}\ }\textbf {\bibinfo {volume} {8}},\ \bibinfo {pages} {119} (\bibinfo {year} {2022})}\BibitemShut {NoStop}%
\bibitem [{\citenamefont {van Dam}(2013)}]{van2013implausible}%
  \BibitemOpen
  \bibfield  {author} {\bibinfo {author} {\bibfnamefont {W.}~\bibnamefont {van Dam}},\ }\href {https://doi.org/10.1007/s11047-012-9353-6} {\bibfield  {journal} {\bibinfo  {journal} {Nat. Comput.}\ }\textbf {\bibinfo {volume} {12}},\ \bibinfo {pages} {9} (\bibinfo {year} {2013})}\BibitemShut {NoStop}%
\bibitem [{\citenamefont {Jones}\ and\ \citenamefont {Masanes}(2005)}]{JM2005PRA}%
  \BibitemOpen
  \bibfield  {author} {\bibinfo {author} {\bibfnamefont {N.~S.}\ \bibnamefont {Jones}}\ and\ \bibinfo {author} {\bibfnamefont {L.}~\bibnamefont {Masanes}},\ }\href {https://doi.org/10.1103/PhysRevA.72.052312} {\bibfield  {journal} {\bibinfo  {journal} {Phys. Rev. A}\ }\textbf {\bibinfo {volume} {72}},\ \bibinfo {pages} {052312} (\bibinfo {year} {2005})}\BibitemShut {NoStop}%
\bibitem [{\citenamefont {Barrett}\ \emph {et~al.}(2005)\citenamefont {Barrett}, \citenamefont {Linden}, \citenamefont {Massar}, \citenamefont {Pironio}, \citenamefont {Popescu},\ and\ \citenamefont {Roberts}}]{Barrett2005}%
  \BibitemOpen
  \bibfield  {author} {\bibinfo {author} {\bibfnamefont {J.}~\bibnamefont {Barrett}}, \bibinfo {author} {\bibfnamefont {N.}~\bibnamefont {Linden}}, \bibinfo {author} {\bibfnamefont {S.}~\bibnamefont {Massar}}, \bibinfo {author} {\bibfnamefont {S.}~\bibnamefont {Pironio}}, \bibinfo {author} {\bibfnamefont {S.}~\bibnamefont {Popescu}},\ and\ \bibinfo {author} {\bibfnamefont {D.}~\bibnamefont {Roberts}},\ }\href {https://doi.org/10.1103/PhysRevA.71.022101} {\bibfield  {journal} {\bibinfo  {journal} {Phys. Rev. A}\ }\textbf {\bibinfo {volume} {71}},\ \bibinfo {pages} {022101} (\bibinfo {year} {2005})}\BibitemShut {NoStop}%
\bibitem [{\citenamefont {Pironio}\ \emph {et~al.}(2011)\citenamefont {Pironio}, \citenamefont {Bancal},\ and\ \citenamefont {Scarani}}]{Pironio2011}%
  \BibitemOpen
  \bibfield  {author} {\bibinfo {author} {\bibfnamefont {S.}~\bibnamefont {Pironio}}, \bibinfo {author} {\bibfnamefont {J.-D.}\ \bibnamefont {Bancal}},\ and\ \bibinfo {author} {\bibfnamefont {V.}~\bibnamefont {Scarani}},\ }\href {https://doi.org/10.1088/1751-8113/44/6/065303} {\bibfield  {journal} {\bibinfo  {journal} {J. Phys. A: Math. Theor.}\ }\textbf {\bibinfo {volume} {44}},\ \bibinfo {pages} {065303} (\bibinfo {year} {2011})}\BibitemShut {NoStop}%
\bibitem [{\citenamefont {Cavalcanti}\ \emph {et~al.}(2010)\citenamefont {Cavalcanti}, \citenamefont {Salles},\ and\ \citenamefont {Scarani}}]{Cavalcanti2010}%
  \BibitemOpen
  \bibfield  {author} {\bibinfo {author} {\bibfnamefont {D.}~\bibnamefont {Cavalcanti}}, \bibinfo {author} {\bibfnamefont {A.}~\bibnamefont {Salles}},\ and\ \bibinfo {author} {\bibfnamefont {V.}~\bibnamefont {Scarani}},\ }\href {https://doi.org/10.1038/ncomms1138} {\bibfield  {journal} {\bibinfo  {journal} {Nat. Commun.}\ }\textbf {\bibinfo {volume} {1}},\ \bibinfo {pages} {136} (\bibinfo {year} {2010})}\BibitemShut {NoStop}%
\bibitem [{\citenamefont {Cabello}(2001)}]{Cabello:2001PRLb}%
  \BibitemOpen
  \bibfield  {author} {\bibinfo {author} {\bibfnamefont {A.}~\bibnamefont {Cabello}},\ }\href {https://doi.org/10.1103/PhysRevLett.87.010403} {\bibfield  {journal} {\bibinfo  {journal} {Phys. Rev. Lett.}\ }\textbf {\bibinfo {volume} {87}},\ \bibinfo {pages} {010403} (\bibinfo {year} {2001})}\BibitemShut {NoStop}%
\bibitem [{\citenamefont {L{\"o}rwald}\ and\ \citenamefont {Reinelt}(2015)}]{lorwald2015panda}%
  \BibitemOpen
  \bibfield  {author} {\bibinfo {author} {\bibfnamefont {S.}~\bibnamefont {L{\"o}rwald}}\ and\ \bibinfo {author} {\bibfnamefont {G.}~\bibnamefont {Reinelt}},\ }\href {https://doi.org/10.1007/s13675-015-0040-0} {\bibfield  {journal} {\bibinfo  {journal} {EURO J. Comput. Optim.}\ }\textbf {\bibinfo {volume} {3}},\ \bibinfo {pages} {297} (\bibinfo {year} {2015})}\BibitemShut {NoStop}%
\bibitem [{Git()}]{GitHub}%
  \BibitemOpen
  \href {https://github.com/sebastiendesignolle/ENS-boxes} {\bibinfo {title} {{ENS}-boxes: extremal non-signaling boxes}},\ \bibinfo {note} {accessed September 30, 2026}\BibitemShut {NoStop}%
\bibitem [{\citenamefont {Ac{\'\i}n}\ \emph {et~al.}(2015)\citenamefont {Ac{\'\i}n}, \citenamefont {Fritz}, \citenamefont {Leverrier},\ and\ \citenamefont {Sainz}}]{acin2015combinatorial}%
  \BibitemOpen
  \bibfield  {author} {\bibinfo {author} {\bibfnamefont {A.}~\bibnamefont {Ac{\'\i}n}}, \bibinfo {author} {\bibfnamefont {T.}~\bibnamefont {Fritz}}, \bibinfo {author} {\bibfnamefont {A.}~\bibnamefont {Leverrier}},\ and\ \bibinfo {author} {\bibfnamefont {A.~B.}\ \bibnamefont {Sainz}},\ }\href {https://doi.org/10.1007/s00220-014-2260-1} {\bibfield  {journal} {\bibinfo  {journal} {Commun. Math. Phys.}\ }\textbf {\bibinfo {volume} {334}},\ \bibinfo {pages} {533} (\bibinfo {year} {2015})}\BibitemShut {NoStop}%
\bibitem [{\citenamefont {Cabello}(2019)}]{cabello2019quantum}%
  \BibitemOpen
  \bibfield  {author} {\bibinfo {author} {\bibfnamefont {A.}~\bibnamefont {Cabello}},\ }\href {https://doi.org/10.1103/PhysRevA.100.032120} {\bibfield  {journal} {\bibinfo  {journal} {Phys. Rev. A}\ }\textbf {\bibinfo {volume} {100}},\ \bibinfo {pages} {032120} (\bibinfo {year} {2019})}\BibitemShut {NoStop}%
\bibitem [{\citenamefont {Chiribella}\ \emph {et~al.}(2020)\citenamefont {Chiribella}, \citenamefont {Cabello}, \citenamefont {Kleinmann},\ and\ \citenamefont {M{\"u}ller}}]{chiribella2020general}%
  \BibitemOpen
  \bibfield  {author} {\bibinfo {author} {\bibfnamefont {G.}~\bibnamefont {Chiribella}}, \bibinfo {author} {\bibfnamefont {A.}~\bibnamefont {Cabello}}, \bibinfo {author} {\bibfnamefont {M.}~\bibnamefont {Kleinmann}},\ and\ \bibinfo {author} {\bibfnamefont {M.~P.}\ \bibnamefont {M{\"u}ller}},\ }\href {https://doi.org/10.1103/physrevresearch.2.042001} {\bibfield  {journal} {\bibinfo  {journal} {Phys. Rev. Res.}\ }\textbf {\bibinfo {volume} {2}},\ \bibinfo {pages} {042001} (\bibinfo {year} {2020})}\BibitemShut {NoStop}%
\bibitem [{\citenamefont {Cabello}\ \emph {et~al.}(2010)\citenamefont {Cabello}, \citenamefont {Severini},\ and\ \citenamefont {Winter}}]{cabello2163contextuality}%
  \BibitemOpen
  \bibfield  {author} {\bibinfo {author} {\bibfnamefont {A.}~\bibnamefont {Cabello}}, \bibinfo {author} {\bibfnamefont {S.}~\bibnamefont {Severini}},\ and\ \bibinfo {author} {\bibfnamefont {A.}~\bibnamefont {Winter}},\ }\href@noop {} {\bibinfo {title} {(non-)contextuality of physical theories as an axiom}} (\bibinfo {year} {2010}),\ \Eprint {https://arxiv.org/abs/1010.2163} {arXiv:1010.2163 [quant-ph]} \BibitemShut {NoStop}%
\bibitem [{\citenamefont {Sainz}\ \emph {et~al.}(2014)\citenamefont {Sainz}, \citenamefont {Fritz}, \citenamefont {Augusiak}, \citenamefont {Brask}, \citenamefont {Chaves}, \citenamefont {Leverrier},\ and\ \citenamefont {Ac{\'\i}n}}]{sainz2014exploring}%
  \BibitemOpen
  \bibfield  {author} {\bibinfo {author} {\bibfnamefont {A.~B.}\ \bibnamefont {Sainz}}, \bibinfo {author} {\bibfnamefont {T.}~\bibnamefont {Fritz}}, \bibinfo {author} {\bibfnamefont {R.}~\bibnamefont {Augusiak}}, \bibinfo {author} {\bibfnamefont {J.~B.}\ \bibnamefont {Brask}}, \bibinfo {author} {\bibfnamefont {R.}~\bibnamefont {Chaves}}, \bibinfo {author} {\bibfnamefont {A.}~\bibnamefont {Leverrier}},\ and\ \bibinfo {author} {\bibfnamefont {A.}~\bibnamefont {Ac{\'\i}n}},\ }\href {https://doi.org/10.1103/PhysRevA.89.032117} {\bibfield  {journal} {\bibinfo  {journal} {Phys. Rev. A}\ }\textbf {\bibinfo {volume} {89}},\ \bibinfo {pages} {032117} (\bibinfo {year} {2014})}\BibitemShut {NoStop}%
\bibitem [{\citenamefont {Choudhary}\ and\ \citenamefont {Barbosa}(2024)}]{choudhary2024exclusivity}%
  \BibitemOpen
  \bibfield  {author} {\bibinfo {author} {\bibfnamefont {R.}~\bibnamefont {Choudhary}}\ and\ \bibinfo {author} {\bibfnamefont {R.~S.}\ \bibnamefont {Barbosa}},\ }\href@noop {} {\bibinfo {title} {Exclusivity principle, {Ramsey} theory, and $n$-cycle {PR} boxes}} (\bibinfo {year} {2024}),\ \Eprint {https://arxiv.org/abs/2411.09773} {arXiv:2411.09773 [quant-ph]} \BibitemShut {NoStop}%
\bibitem [{\citenamefont {Designolle}\ \emph {et~al.}(2023)\citenamefont {Designolle}, \citenamefont {Iommazzo}, \citenamefont {Besan{\c{c}}on}, \citenamefont {Knebel}, \citenamefont {Gel{\ss}},\ and\ \citenamefont {Pokutta}}]{designolle2023improved}%
  \BibitemOpen
  \bibfield  {author} {\bibinfo {author} {\bibfnamefont {S.}~\bibnamefont {Designolle}}, \bibinfo {author} {\bibfnamefont {G.}~\bibnamefont {Iommazzo}}, \bibinfo {author} {\bibfnamefont {M.}~\bibnamefont {Besan{\c{c}}on}}, \bibinfo {author} {\bibfnamefont {S.}~\bibnamefont {Knebel}}, \bibinfo {author} {\bibfnamefont {P.}~\bibnamefont {Gel{\ss}}},\ and\ \bibinfo {author} {\bibfnamefont {S.}~\bibnamefont {Pokutta}},\ }\href {https://doi.org/10.1103/PhysRevResearch.5.043059} {\bibfield  {journal} {\bibinfo  {journal} {Phys. Rev. Res.}\ }\textbf {\bibinfo {volume} {5}},\ \bibinfo {pages} {043059} (\bibinfo {year} {2023})}\BibitemShut {NoStop}%
\bibitem [{\citenamefont {Braun}\ \emph {et~al.}(2022)\citenamefont {Braun}, \citenamefont {Carderera}, \citenamefont {Combettes}, \citenamefont {Hassani}, \citenamefont {Karbasi}, \citenamefont {Mokhtari},\ and\ \citenamefont {Pokutta}}]{braun2022conditional}%
  \BibitemOpen
  \bibfield  {author} {\bibinfo {author} {\bibfnamefont {G.}~\bibnamefont {Braun}}, \bibinfo {author} {\bibfnamefont {A.}~\bibnamefont {Carderera}}, \bibinfo {author} {\bibfnamefont {C.~W.}\ \bibnamefont {Combettes}}, \bibinfo {author} {\bibfnamefont {H.}~\bibnamefont {Hassani}}, \bibinfo {author} {\bibfnamefont {A.}~\bibnamefont {Karbasi}}, \bibinfo {author} {\bibfnamefont {A.}~\bibnamefont {Mokhtari}},\ and\ \bibinfo {author} {\bibfnamefont {S.}~\bibnamefont {Pokutta}},\ }\href@noop {} {\bibinfo {title} {Conditional gradient methods}} (\bibinfo {year} {2022}),\ \Eprint {https://arxiv.org/abs/2211.14103} {arXiv:2211.14103 [math.OC]} \BibitemShut {NoStop}%
\bibitem [{\citenamefont {Brunner}\ \emph {et~al.}(2006)\citenamefont {Brunner}, \citenamefont {Scarani},\ and\ \citenamefont {Gisin}}]{Brunner2006}%
  \BibitemOpen
  \bibfield  {author} {\bibinfo {author} {\bibfnamefont {N.}~\bibnamefont {Brunner}}, \bibinfo {author} {\bibfnamefont {V.}~\bibnamefont {Scarani}},\ and\ \bibinfo {author} {\bibfnamefont {N.}~\bibnamefont {Gisin}},\ }\href {https://doi.org/10.1063/1.2352857} {\bibfield  {journal} {\bibinfo  {journal} {J. Math. Phys.}\ }\textbf {\bibinfo {volume} {47}},\ \bibinfo {pages} {112101} (\bibinfo {year} {2006})}\BibitemShut {NoStop}%
\bibitem [{\citenamefont {Wolfe}\ and\ \citenamefont {Yelin}(2012)}]{Wolfe2012}%
  \BibitemOpen
  \bibfield  {author} {\bibinfo {author} {\bibfnamefont {E.}~\bibnamefont {Wolfe}}\ and\ \bibinfo {author} {\bibfnamefont {S.~F.}\ \bibnamefont {Yelin}},\ }\href {https://doi.org/10.1103/PhysRevA.86.012123} {\bibfield  {journal} {\bibinfo  {journal} {Phys. Rev. A}\ }\textbf {\bibinfo {volume} {86}},\ \bibinfo {pages} {012123} (\bibinfo {year} {2012})}\BibitemShut {NoStop}%
\bibitem [{\citenamefont {Cope}\ and\ \citenamefont {Colbeck}(2019)}]{Cope2019}%
  \BibitemOpen
  \bibfield  {author} {\bibinfo {author} {\bibfnamefont {T.}~\bibnamefont {Cope}}\ and\ \bibinfo {author} {\bibfnamefont {R.}~\bibnamefont {Colbeck}},\ }\href {https://doi.org/10.1103/PhysRevA.100.022114} {\bibfield  {journal} {\bibinfo  {journal} {Phys. Rev. A}\ }\textbf {\bibinfo {volume} {100}},\ \bibinfo {pages} {022114} (\bibinfo {year} {2019})}\BibitemShut {NoStop}%
\bibitem [{\citenamefont {Ramanathan}(2021)}]{Ravi2021PRR}%
  \BibitemOpen
  \bibfield  {author} {\bibinfo {author} {\bibfnamefont {R.}~\bibnamefont {Ramanathan}},\ }\href {https://doi.org/10.1103/PhysRevResearch.3.033100} {\bibfield  {journal} {\bibinfo  {journal} {Phys. Rev. Res.}\ }\textbf {\bibinfo {volume} {3}},\ \bibinfo {pages} {033100} (\bibinfo {year} {2021})}\BibitemShut {NoStop}%
\bibitem [{\citenamefont {Liu}\ \emph {et~al.}(2024{\natexlab{b}})\citenamefont {Liu}, \citenamefont {Chung},\ and\ \citenamefont {Ramanathan}}]{Liu2023}%
  \BibitemOpen
  \bibfield  {author} {\bibinfo {author} {\bibfnamefont {Y.}~\bibnamefont {Liu}}, \bibinfo {author} {\bibfnamefont {H.~Y.}\ \bibnamefont {Chung}},\ and\ \bibinfo {author} {\bibfnamefont {R.}~\bibnamefont {Ramanathan}},\ }\href {https://doi.org/10.22331/q-2024-10-02-1489} {\bibfield  {journal} {\bibinfo  {journal} {Quantum}\ }\textbf {\bibinfo {volume} {8}},\ \bibinfo {pages} {1489} (\bibinfo {year} {2024}{\natexlab{b}})}\BibitemShut {NoStop}%
\bibitem [{\citenamefont {Masanes}\ \emph {et~al.}(2014)\citenamefont {Masanes}, \citenamefont {Renner}, \citenamefont {Christandl}, \citenamefont {Winter},\ and\ \citenamefont {Barrett}}]{Masanes2014}%
  \BibitemOpen
  \bibfield  {author} {\bibinfo {author} {\bibfnamefont {L.}~\bibnamefont {Masanes}}, \bibinfo {author} {\bibfnamefont {R.}~\bibnamefont {Renner}}, \bibinfo {author} {\bibfnamefont {M.}~\bibnamefont {Christandl}}, \bibinfo {author} {\bibfnamefont {A.}~\bibnamefont {Winter}},\ and\ \bibinfo {author} {\bibfnamefont {J.}~\bibnamefont {Barrett}},\ }\href {https://doi.org/10.1109/tit.2014.2329417} {\bibfield  {journal} {\bibinfo  {journal} {IEEE Trans. Inf. Theory}\ }\textbf {\bibinfo {volume} {60}},\ \bibinfo {pages} {4973} (\bibinfo {year} {2014})}\BibitemShut {NoStop}%
\end{thebibliography}%
\vspace{-10pt}

\appendix
\setcounter{secnumdepth}{1}

\section{Illustrative example of candidate ENS boxes}\label{sec:exampleCandidate}

For $X=Y=4$, $A=3$, and $g=h=2$, \cref{eq:1} becomes
\begin{equation}
  \begin{pmatrix}
    S & S      & L & L \\
    S & T_{22} & L & L \\
    K & K      & M & M \\
    K & K      & M & M
  \end{pmatrix},
\end{equation}
which reads, more explicitly,
\begin{equation}
  \resizebox{!}{7em}{$
  \left(\begin{array}{ccc|ccc|ccc|ccc}
    \frac13 &         &         & \frac13 &         &         & \frac13 & \hphantom{1} & \hphantom{1} & \frac13 & \hphantom{1} & \hphantom{1} \\
            & \frac13 &         &         & \frac13 &         & \frac13 &              &              & \frac13 &              &              \\
            &         & \frac13 &         &         & \frac13 & \frac13 &              &              & \frac13 &              &              \\
    \hline
    \frac13 &         &         &         & \frac13 &         & \frac13 &              &              & \frac13 &              &              \\
            & \frac13 &         &         &         & \frac13 & \frac13 &              &              & \frac13 &              &              \\
            &         & \frac13 & \frac13 &         &         & \frac13 &              &              & \frac13 &              &              \\
    \hline
    \frac13 & \frac13 & \frac13 & \frac13 & \frac13 & \frac13 & 1       &              &              & 1       &              &              \\
            &         &         &         &         &         &         &              &              &         &              &              \\
            &         &         &         &         &         &         &              &              &         &              &              \\
    \hline
    \frac13 & \frac13 & \frac13 & \frac13 & \frac13 & \frac13 & 1       &              &              & 1       &              &              \\
            &         &         &         &         &         &         &              &              &         &              &              \\
            &         &         &         &         &         &         &              &              &         &              &              \\
  \end{array}\right).
  $}
\end{equation}

\section{ENS boxes in small Bell scenarios}\label{app:small-scenarios}

We analyzed and classified all the ENS boxes, full-output and partial-output, in the scenarios $(2,3,3,2)$, $(3,3,3,2)$, and $(2,3,3,3)$.
We follow the terminology introduced in Ref.~\cite{Barrett2005}.
Thus, we refer to correlations with all marginals, $p(a|x)$ and $p(b|y)$, greater than zero as ``full-output'' correlations.
Correlations for which at least one marginal is zero are called ``partial-output.''

\subsection{Scenario (2,3,3,2)}

In this case there are four ENS boxes.
One of them is local and, up to relabelings, generates all the vertices of the local polytope for this scenario.
The remaining three vertices are nonlocal.
None of these three classes is full-output.

\subsection{Scenario (3,3,3,2)}

In this case there are 16 classes of nonlocal ENS boxes; of these, only two are full-output: $V^1_{(3,3,3,2)}$ and $V^2_{(3,3,3,2)}$.

\begin{equation}
  \label{eq:box-1}
  V^1_{(3,3,3,2)}=\frac{1}{4}\times\left(
  \begin{array}{cc|cc|cc}
      & 1 &   & 1 &   & 1 \\
      & 1 & 1 &   & 1 &   \\
    2 &   &   & 2 &   & 2 \\
    \hline
      & 2 &   & 2 &   & 2 \\
    1 &   &   & 1 & 1 &   \\
    1 &   & 1 &   &   & 1 \\
    \hline
      & 1 &   & 1 & 1 &   \\
      & 1 & 1 &   &   & 1 \\
    2 &   &   & 2 &   & 2
  \end{array}
  \right),
\end{equation}

\begin{equation}
  \label{eq:box-2}
  V^2_{(3,3,3,2)}=\frac{1}{3}\times\left(
  \begin{array}{cc|cc|cc}
      & 1 &   & 1 &   & 1 \\
      & 1 &   & 1 & 1 &   \\
    1 &   & 1 &   &   & 1 \\
    \hline
      & 1 &   & 1 &   & 1 \\
      & 1 & 1 &   &   & 1 \\
    1 &   &   & 1 & 1 &   \\
    \hline
      & 1 &   & 1 & 1 &   \\
      & 1 & 1 &   &   & 1 \\
    1 &   &   & 1 &   & 1
  \end{array}
  \right).
\end{equation}

\subsection{Scenario (2,3,3,3)}

In this scenario there are 22 classes of nonlocal NS vertices; of these, six are full-output vertices.
Three examples are shown next.
These have 34, 35, and 36 zeros.

\begin{equation}
  \label{eq:box-3}
  V^{34}_{(2,3,3,3)}=\frac{1}{4}\times\left(
  \begin{array}{ccc|ccc|ccc}
      &   & 2 &   &   & 2 &   &   & 2 \\
      & 1 &   &   & 1 &   &   & 1 &   \\
    1 &   &   & 1 &   &   & 1 &   &   \\
    \hline
      &   & 1 &   &   & 1 &   & 1 &   \\
      &   & 1 &   & 1 &   & 1 &   &   \\
    1 & 1 &   & 1 &   & 1 &   &   & 2
  \end{array}
  \right),
\end{equation}

\begin{equation}
  \label{eq:box-4}
  V^{35}_{(2,3,3,3)}=\frac{1}{4}\times\left(
  \begin{array}{ccc|ccc|ccc}
      &   & 1 &   &   & 1 &   &   & 1 \\
      & 1 &   &   & 1 &   &   &   & 1 \\
    2 &   &   & 2 &   &   & 1 & 1 &   \\
    \hline
      &   & 1 &   & 1 &   &   & 1 &   \\
      & 1 &   &   &   & 1 & 1 &   &   \\
    2 &   &   & 2 &   &   &   &   & 2
  \end{array}
  \right),
\end{equation}

\begin{equation}
  \label{eq:box-5}
  V^{36}_{(2,3,3,3)}=\frac{1}{3}\times\left(
  \begin{array}{ccc|ccc|ccc}
      &   & 1 &   &   & 1 &   &   & 1 \\
      & 1 &   &   & 1 &   &   & 1 &   \\
    1 &   &   & 1 &   &   & 1 &   &   \\
    \hline
      &   & 1 &   &   & 1 &   & 1 &   \\
      & 1 &   &   & 1 &   & 1 &   &   \\
    1 &   &   & 1 &   &   &   &   & 1
  \end{array}
  \right).
\end{equation}

\section{A family of full-output vertices in \texorpdfstring{$(X,Y,d,d)$}{(X,Y,d,d)}}

\begin{figure}[!htbp]
  \makebox[\linewidth][c]{%
  \begin{minipage}{0.48\linewidth}
    \centering
    \begin{tikzpicture}[scale=0.3, baseline={(current bounding box.south)}]
      \draw[step=1cm,gray,very thin] (2,2) grid (10,10);
      \foreach \x/\y in {
      2/2,2/3,2/4,2/5,2/6,2/7,2/8,2/9,
      3/2,3/3,3/4,3/5,3/6,3/7,
      4/2,4/3,4/4,4/5,
      5/2,5/3,5/4,5/5,
      6/2,6/3,
      7/2,7/3,
      8/2,9/2
      }
      \fill[gray!80] (\x,\y) rectangle+(1,1);
      \foreach \x/\y in {3/8,4/6,6/4,8/3}
      \fill[gray!40] (\x,\y) rectangle+(1,1);
      \foreach \x/\y in {2/2,3/3,4/4,5/5,6/6,7/7,8/8,9/9}
      \fill[pattern=north east lines] (\x,\y) rectangle+(1,1);
      \foreach \x/\y in {2/2}
      \fill[pattern=north west lines] (\x,\y) rectangle+(1,1);
      \draw[->, thick] (2,2) -- (10.5,2) node[right] {$X$};
      \draw[->, thick] (2,2) -- (2,10.5) node[above] {$Y$};
      \foreach \x in {2,...,9} \draw (\x+0.5,2.1) -- (\x+0.5,1.9) node[below,scale=0.6] {\x};
      \foreach \y in {2,...,9} \draw (2.1,\y+0.5) -- (1.9,\y+0.5) node[left,scale=0.6] {\y};
      \node[below=10pt, align=center] at (6,2) {$(A,B)=(2,2)$};
    \end{tikzpicture}
  \end{minipage}
  \hfill
  \begin{minipage}{0.48\linewidth}
    \centering
    \begin{tikzpicture}[scale=0.3, baseline={(current bounding box.south)}]
      \draw[step=1cm,gray,very thin] (2,2) grid (10,10);
      \foreach \x/\y in {2/2,2/3,2/4,2/5,2/6,2/7,3/2,3/3,3/4,4/2,4/3,5/2}
      \fill[gray!80] (\x,\y) rectangle+(1,1);
      \draw[->, thick] (2,2) -- (10.5,2) node[right] {$X$};
      \draw[->, thick] (2,2) -- (2,10.5) node[above] {$Y$};
      \foreach \x in {2,...,9} \draw (\x+0.5,2.1) -- (\x+0.5,1.9) node[below,scale=0.6] {\x};
      \foreach \y in {2,...,9} \draw (2.1,\y+0.5) -- (1.9,\y+0.5) node[left,scale=0.6] {\y};
      \node[below=10pt, align=center] at (6,2) {$(A,B)=(2,3)$};
    \end{tikzpicture}
  \end{minipage}
  }

  \vspace{3ex}

  \makebox[\linewidth][c]{%
  \begin{minipage}{0.48\linewidth}
    \centering
    \begin{tikzpicture}[scale=0.3, baseline={(current bounding box.south)}]
      \draw[step=1cm,gray,very thin] (2,2) grid (10,10);
      \foreach \x/\y in {2/2,2/3,2/4,2/5,3/2,3/3,4/2}
      \fill[gray!80] (\x,\y) rectangle+(1,1);
      \foreach \x/\y in {5/2}
      \fill[gray!40] (\x,\y) rectangle+(1,1);
      \draw[->, thick] (2,2) -- (10.5,2) node[right] {$X$};
      \draw[->, thick] (2,2) -- (2,10.5) node[above] {$Y$};
      \foreach \x in {2,...,9} \draw (\x+0.5,2.1) -- (\x+0.5,1.9) node[below,scale=0.6] {\x};
      \foreach \y in {2,...,9} \draw (2.1,\y+0.5) -- (1.9,\y+0.5) node[left,scale=0.6] {\y};
      \node[below=10pt, align=center] at (6,2) {$(A,B)=(2,4)$};
    \end{tikzpicture}
  \end{minipage}
  \hfill
  \begin{minipage}{0.48\linewidth}
    \centering
    \begin{tikzpicture}[scale=0.3, baseline={(current bounding box.south)}]
      \draw[step=1cm,gray,very thin] (2,2) grid (10,10);
      \foreach \x/\y in {2/2,2/3,2/4,3/2,4/2}
      \fill[gray!80] (\x,\y) rectangle+(1,1);
      \foreach \x/\y in {3/3}
      \fill[gray!40] (\x,\y) rectangle+(1,1);
      \foreach \x/\y in {2/2}
      \fill[pattern=north west lines] (\x,\y) rectangle+(1,1);
      \draw[->, thick] (2,2) -- (10.5,2) node[right] {$X$};
      \draw[->, thick] (2,2) -- (2,10.5) node[above] {$Y$};
      \foreach \x in {2,...,9} \draw (\x+0.5,2.1) -- (\x+0.5,1.9) node[below,scale=0.6] {\x};
      \foreach \y in {2,...,9} \draw (2.1,\y+0.5) -- (1.9,\y+0.5) node[left,scale=0.6] {\y};
      \node[below=10pt, align=center] at (6,2) {$(A,B)=(3,3)$};
    \end{tikzpicture}
  \end{minipage}
  }

  \caption{
  Illustration showing the bipartite Bell scenarios for which complete or conjecturally complete lists of ENS boxes have been obtained.
  Dark gray cells indicate scenarios solved in this work, while light gray ones correspond to scenarios for which we only conjecture our list of extreme points to be complete.
  The scenarios $(2,2,d,d)$ and $(d,d,2,2)$ were respectively solved in~\cite{Barrett2005,JM2005PRA} and are shown with hatch patterns.
  }
  \label{fig:scenarios1}
\end{figure}
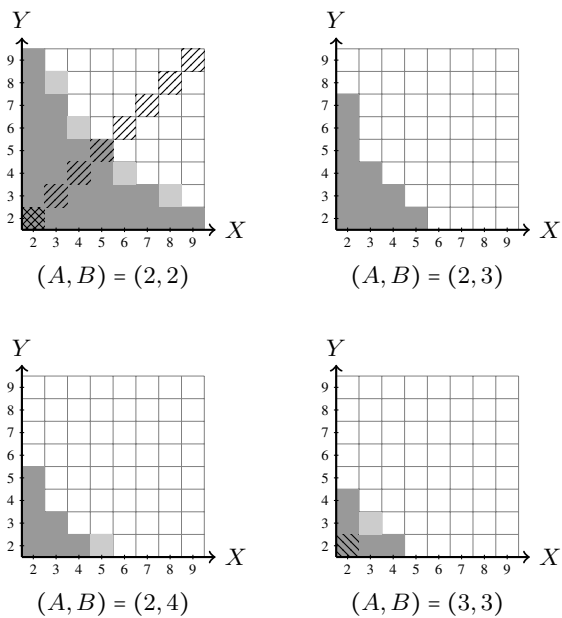

In \cref{fig:scenarios1,tab:scenarios2}, we summarize the scenarios for which we have obtained complete or partial lists of ENS boxes.

\begin{table}[!htbp]
  \centering
  \begin{minipage}{\linewidth}
    \centering
    \subfloat{
    \begin{tabular}{c @{\hspace{12pt}} c @{\hspace{12pt}} c}
      $(X,Y,A,B)$ & \#L1 & Ref.               \\ \hline
      (2,2,3,4)   & 2    & This work          \\
      (2,2,4,4)   & 3    & \cite{Barrett2005} \\
      (2,2,5,5)   & 4    & \cite{Barrett2005} \\
      (2,3,3,4)   & 28   & This work          \\
      (2,3,4,3)   & 104  & This work          \\
    \end{tabular}
    }
  \end{minipage}%
  \caption{
  Number of classes of nonlocal ENS boxes.
  For each scenario, we provide a complete list of classes.
  The total number of classes is the number of nonlocal ones (column \#L1) plus one.
  }
  \label{tab:scenarios2}
\end{table}

It is clear that a wider variety of boxes appears when the number of inputs and outputs is greater than two, as exemplified by the scenario (2,3,3,3) in the previous section.
Thus, a complete characterization for arbitrary scenarios is an increasingly complex task.
However, it is still possible to make progress on this problem by providing partial characterizations in scenarios with particular properties.
For instance, we derive a family of full-output ENS boxes in scenarios where the number of zeros in the table of correlations $Z$ is equal to $\mathrm{dim}(\mathcal{NS})$, the dimension of the corresponding NS polytope.

In any scenario with $X\geq2$ and $Y\geq2$, the inequality $Z\leq\mathrm{dim}(\mathcal{NS})$ holds.
Therefore, there are NS vertices with the maximum number of zeros.
In fact, we can follow the same argument as in~\cite{Barrett2005} and find that there is a family of vertices of the following form
\begin{equation}\label{eq.NSdd}
  p(a,b|x,y)=\left(
  \begin{array}{c|c|c|c|c}
    D      & D      & D      & \dots & D      \\
    \hline
    D      & P^{k}  & P      & \dots & P      \\
    \hline
    D      & P      & P      & \dots & P      \\
    \hline
    \vdots & \vdots & \vdots &       & \vdots \\
    \hline
    D      & P      & P      & \dots & P
  \end{array}
  \right),
\end{equation}
where $D$ is a diagonal matrix, $P^{k}$ is a permutation of order $k$, and $P$ is a permutation of any order.
In addition, every probability in the table is equal to $1/d$.

One example of this type of vertex is, after relabelings, $V^{36}_{(2,3,3,3)}$.
The construction in \cref{eq.NSdd} is a natural generalization of the boxes derived in~\cite{Barrett2005} and~\cite{JM2005PRA}.

\section{The magic square correlations}\label{app:C}
The result in~\cite{Ramanathan:2010PRL} implies that the correlations that win the magic square game are not extremal points in $\mathcal{NS}$.
Thus, it is possible to write them as a convex combination of ENS boxes.
This was explicitly shown in the main text.
Here we review the correlations that win the magic square game and discuss some of their properties.

First, the original definition of the game specifies conditions for filling a $3\times3$ table.
These conditions can always be fulfilled using the observables in the Peres--Mermin magic square
\begin{equation}\label{tab.PMsquare}
  \begin{tabular}{|c|c|c|}
    \hline
    $\sigma_z\otimes\mathds{1}$ & $\mathds{1}\otimes\sigma_z$ & $\sigma_z\otimes\sigma_z$ \\
    \hline
    $\mathds{1}\otimes\sigma_x$ & $\sigma_x\otimes\mathds{1}$ & $\sigma_x\otimes\sigma_x$ \\
    \hline
    $\sigma_z\otimes\sigma_x$   & $\sigma_x\otimes\sigma_z$   & $\sigma_y\otimes\sigma_y$ \\
    \hline
  \end{tabular}
\end{equation}
At first glance, this is a game with three inputs and eight outcomes.
However, the third outcome for each measurement is completely determined from the conditions of the game.
Therefore, this can be recast as a three-input, four-outcome game in the (3,3,4,4) scenario.
The specific measurements can be obtained by simultaneously diagonalizing the three operators in each row and column of \cref{tab.PMsquare}.
The diagonalization yields 24 projectors, which can be arranged in six bases and distributed to Alice and Bob as follows:

\begin{subequations}
  \label{eq.PMproj}
  \begin{align}
      & A^{0|0}=(1,0,0,0),   & B^{0|0} & =(1,1,0,0),  \\
      & A^{1|0}=(0,1,0,0),   & B^{1|0} & =(1,-1,0,0), \\
      & A^{2|0}=(0,0,1,0),   & B^{2|0} & =(0,0,1,1),  \\
      & A^{3|0}=(0,0,0,1),   & B^{3|0} & =(0,0,1,-1), \\
      & A^{0|1}=(1,1,1,1),   & B^{0|1} & =(1,0,1,0),  \\
      & A^{1|1}=(1,1,-1,-1), & B^{1|1} & =(1,0,-1,0), \\
      & A^{2|1}=(1,-1,1,-1), & B^{2|1} & =(0,1,0,1),  \\
      & A^{3|1}=(1,-1,-1,1), & B^{3|1} & =(0,1,0,-1), \\
      & A^{0|2}=(1,1,1,-1),  & B^{0|2} & =(1,0,0,-1), \\
      & A^{1|2}=(1,1,-1,1),  & B^{1|2} & =(1,0,0,1),  \\
      & A^{2|2}=(1,-1,1,1),  & B^{2|2} & =(0,1,-1,0), \\
      & A^{3|2}=(-1,1,1,1),  & B^{3|2} & =(0,1,1,0).
  \end{align}
\end{subequations}
These projectors, together with the maximally entangled state $\ket{\psi}=\frac12 \sum_i \ket{ii}$, produce the magic square correlations~\cite{Cabello:2001PRLb}:

\begin{equation}
  \vec{p}_{MS}=
  \frac{1}{8}\times
  \resizebox{!}{5em}{$
  \left(
  \begin{array}{cccc|cccc|cccc}
    1 & 1 &   &   & 1 & 1 &   &   & 1 & 1 &   &   \\
    1 & 1 &   &   &   &   & 1 & 1 &   &   & 1 & 1 \\
      &   & 1 & 1 & 1 & 1 &   &   &   &   & 1 & 1 \\
      &   & 1 & 1 &   &   & 1 & 1 & 1 & 1 &   &   \\
    \hline
    1 &   & 1 &   & 1 &   & 1 &   &   & 1 &   & 1 \\
    1 &   & 1 &   &   & 1 &   & 1 & 1 &   & 1 &   \\
      & 1 &   & 1 & 1 &   & 1 &   & 1 &   & 1 &   \\
      & 1 &   & 1 &   & 1 &   & 1 &   & 1 &   & 1 \\
    \hline
    1 &   &   & 1 & 1 &   &   & 1 & 1 &   &   & 1 \\
    1 &   &   & 1 &   & 1 & 1 &   &   & 1 & 1 &   \\
      & 1 & 1 &   & 1 &   &   & 1 &   & 1 & 1 &   \\
      & 1 & 1 &   &   & 1 & 1 &   & 1 &   &   & 1
  \end{array}
  \right)
  $}.
\end{equation}
Using the magic square correlations $\vec{p}_{MS}$, it is straightforward to find a Bell inequality for which they attain the algebraic maximum of 9:

\begin{equation}
  B_{MS}=
  \resizebox{!}{5em}{$
  \left(
  \begin{array}{cccc|cccc|cccc}
    1 & 1 &   &   & 1 & 1 &   &   & 1 & 1 &   &   \\
    1 & 1 &   &   &   &   & 1 & 1 &   &   & 1 & 1 \\
      &   & 1 & 1 & 1 & 1 &   &   &   &   & 1 & 1 \\
      &   & 1 & 1 &   &   & 1 & 1 & 1 & 1 &   &   \\
    \hline
    1 &   & 1 &   & 1 &   & 1 &   &   & 1 &   & 1 \\
    1 &   & 1 &   &   & 1 &   & 1 & 1 &   & 1 &   \\
      & 1 &   & 1 & 1 &   & 1 &   & 1 &   & 1 &   \\
      & 1 &   & 1 &   & 1 &   & 1 &   & 1 &   & 1 \\
    \hline
    1 &   &   & 1 & 1 &   &   & 1 & 1 &   &   & 1 \\
    1 &   &   & 1 &   & 1 & 1 &   &   & 1 & 1 &   \\
      & 1 & 1 &   & 1 &   &   & 1 &   & 1 & 1 &   \\
      & 1 & 1 &   &   & 1 & 1 &   & 1 &   &   & 1
  \end{array}
  \right)
  $}.
\end{equation}
The magic square game is obtained simply by normalizing this inequality using the factor $1/9$.

Here we prove that $\vec{p}_1$ is an ENS box; the proof for $\vec{p}_2$ follows the same logic.
Consider $\vec{v}=v(a,b|x,y)$, such that $(\vec{p}_1\pm\vec{v})=p(a,b|x,y)\pm v(a,b|x,y)$.
We start by focusing on the set of inputs $x=y=1$.
Given that each component of $(\vec{p}_1\pm\vec{v})$ is a probability, we can easily see that $v(a,b|1,1)=0$ for $a\neq b$.
Applying the same constraint to the remaining input pairs $(x,y)$ sets several components of $\vec{v}$ to zero.
Now, $(\vec{p}_1\pm\vec{v})(a,b|x=1,y=1)$ has the remaining coefficients:

\begin{equation}\nonumber
  \resizebox{\linewidth}{!}{$
  \begin{split}
    (\vec{p_1} \pm \vec{v})(a,b) &=
    \left(
      \begin{array}{cccc}
        1/4\pm c_1 & 0  & 0  & 0 \\
        0 & 1/4 \pm c_2 & 0 & 0  \\
        0 & 0 & 1/4 \pm c_3 & 0  \\
        0 & 0 & 0 & 1/4 \pm c_4  \\
      \end{array}
    \right)\\
    &= \left(
      \begin{array}{cccc}
        1/4&   &   &     \\
           & 1/4 &   &   \\
           &   & 1/4 &   \\
           &   &   & 1/4 \\
      \end{array}
    \right) \pm \left(
      \begin{array}{cccc}
        c_1 &    &    &   \\
            & c_2 &   &   \\
            &   & c_3 &   \\
            &   &   & c_4 \\
      \end{array}
    \right)
  \end{split}
  $}
\end{equation}
where $c_i=v(a=i,b=i|x=1,y=1)$.
Normalization, $\sum_{a,b} p(a,b|x,y)=1$, gives the constraint $\sum_i c_i=0$.
Now we apply the $\mathcal{NS}$ condition, which is the final constraint that sets $\vec{v}=0$.
This step can be performed numerically, but for illustrative purposes we show it in two steps.
First, we use as a reference the input $(x=1,y=1)$, and we define the coefficients in $(x=1,y=2)$ and $(x=1,y=3)$ by demanding that the marginal probabilities should be the same $P^A(a|x=1)=\sum_b P(a,b|x=1,y'),\quad\text{for all }(a,y')$.
Next, we use the inputs with $x=1$ to fill $P^B(b|y=1)$, $P^B(b|y=2)$, and $P^B(b|y=3)$ using the condition of the marginals.
Therefore, $(\vec{p}_1\pm\vec{v})$ looks like

\begin{equation}\nonumber
  (\vec{p}_1\pm\vec{v})=\frac{1}{4}\times
  \resizebox{!}{4.5em}{$
  \left(
  \begin{array}{cccc|cccc|cccc}
    1 &   &   &   & 1 &   &   &   & 1 &   &   &   \\
      & 1 &   &   &   &   & 1 &   &   &   & 1 &   \\
      &   & 1 &   &   & 1 &   &   &   &   &   & 1 \\
      &   &   & 1 &   &   &   & 1 &   & 1 &   &   \\
    \hline
    1 &   &   &   & 1 &   &   &   &   & 1 &   &   \\
      &   & 1 &   &   & 1 &   &   & 1 &   &   &   \\
      & 1 &   &   &   &   & 1 &   &   &   & 1 &   \\
      &   &   & 1 &   &   &   & 1 &   &   &   & 1 \\
    \hline
    1 &   &   &   & 1 &   &   &   & 1 &   &   &   \\
      &   &   & 1 &   & 1 &   &   &   & 1 &   &   \\
      & 1 &   &   &   &   &   & 1 &   &   & 1 &   \\
      &   & 1 &   &   &   & 1 &   &   &   &   & 1
  \end{array}
  \right)
  $}\pm
  \resizebox{!}{4.5em}{$
  \left(
  \begin{array}{cccc|cccc|cccc}
    c_1 &     &     &     & c_1 &     &     &     & c_1 &     &     &     \\
        & c_2 &     &     &     &     & c_2 &     &     &     & c_2 &     \\
        &     & c_3 &     &     & c_3 &     &     &     &     &     & c_3 \\
        &     &     & c_4 &     &     &     & c_4 &     & c_4 &     &     \\
    \hline
    c_1 &     &     &     & c_1 &     &     &     &     & c_4 &     &     \\
        &     & c_3 &     &     & c_3 &     &     & c_1 &     &     &     \\
        & c_2 &     &     &     &     & c_2 &     &     &     & c_2 &     \\
        &     &     & c_4 &     &     &     & c_4 &     &     &     & c_3 \\
    \hline
    c_1 &     &     &     & c_1 &     &     &     & c_1 &     &     &     \\
        &     &     & c_4 &     & c_3 &     &     &     & c_4 &     &     \\
        & c_2 &     &     &     &     &     & c_4 &     &     & c_2 &     \\
        &     & c_3 &     &     &     & c_2 &     &     &     &     & c_3
  \end{array}
  \right),
  $}
\end{equation}
Finally, using the consistency conditions for $P^A(a|x=2)$ and $P^A(a|x=3)$, we see that this can only be achieved if
\begin{equation}
  c_1=c_2=c_3=c_4.
\end{equation}
Together with normalization, this condition shows that $c_i=0$ and hence $\vec{v}=0$ is the only solution.
Therefore, $\vec{p}_1$ is an ENS box.

\section{LO\texorpdfstring{$^2$}{2} violation analysis examples}
\label{app:LO2}

For the extremal box given in \cref{eq:box-2}, each event in the support has probability $1/3$.
Thus every joint event in the support of two independent copies has probability $1/9$.
Hence, the condition for a clique of size $l$ to violate the inequality is
\begin{equation}
  \frac{l}{9}>1.
\end{equation}
Therefore, the violating clique will be of size 10 or more.
Our code found a $K_{10}$ that violates the corresponding LO$^2$ inequality.
The set of these 10 locally orthogonal events is $\{2021|0002,\allowbreak 2001|0002,\allowbreak 1020|1112,\allowbreak 1120|0120,\allowbreak 1111|0010,\allowbreak 1101|0010,\allowbreak 2000|1022,\allowbreak 0101|0010,\allowbreak 0120|0120,\allowbreak 0111|0010\}$.
The sum of the probabilities of these pairwise exclusive events is $10/9>1$.

We now move to the extremal boxes of scenario (2,3,3,3) and start with the box given in \cref{eq:box-3}.
Running our code for this box, we obtain a $K_{12}$ that violates the corresponding LO$^2$ inequality.
The set of these 12 locally orthogonal events is $\{0212|0110,\allowbreak 0201|1012,\allowbreak 1002|1211,\allowbreak 1111|0001,\allowbreak 2012|0110,\allowbreak 1120|0202,\allowbreak 1102|0202,\allowbreak 2202|1211,\allowbreak 2222|1211,\allowbreak 2220|1211,\allowbreak 2020|0210,\allowbreak 2021|0210\}$.
The sum of the probabilities of these pairwise exclusive events is $17/16>1$.

For the extremal box in \cref{eq:box-4}, we also get a $K_{12}$ that violates the corresponding LO$^2$ inequality.
The set of these 12 locally orthogonal events is $\{0212|0111,\allowbreak 0211|0001,\allowbreak 0120|1211,\allowbreak 2022|1012,\allowbreak 1020|1211,\allowbreak 2011|0010,\allowbreak 2001|0211,\allowbreak 2101|0211,\allowbreak 1221|0202,\allowbreak 1220|0202,\allowbreak 1102|0001,\allowbreak 1111|0001\}$.
The sum of the probabilities of these pairwise exclusive events is $9/8>1$.

For the extremal box in \cref{eq:box-5}, a clique of size 10 or more will be needed.
We found a $K_{10}$, and the corresponding set of 10 mutually orthogonal events is $\{0211|1101,\allowbreak 0220|1101,\allowbreak 1102|0200,\allowbreak 0202|0200,\allowbreak 2002|0110,\allowbreak 2020|1010,\allowbreak 2011|1010,\allowbreak 1022|1212,\allowbreak 1111|1102,\allowbreak 1120|1102\}$.
The sum of the probabilities of these pairwise exclusive events is $10/9>1$.

\section{Communication complexity}
\label{app:dit}

For scenarios in which we can obtain complete lists of classes of extremal NS points, we can test how much communication is required to reproduce the correlations of these different points.
In \cref{tab:comm}, we give the results of our extensive computations, based on which we obtain violations of LHV+$d$it for $d\leq5$, the highest communication complexity being achieved in the scenario $(6,4,2,2)$.

\begin{table*}[t]
  \small
  \renewcommand{\arraystretch}{1.15}
  \centering
  \begin{minipage}{0.4\textwidth}
    \centering
    \subfloat[\label{tab:comm22}]{
    \begin{tabular}{|cc|cc|cc|cc|cc|cc|}
      \multicolumn{12}{c}{$A=2$ and $B=2$} \\ \hline
      $X$ & $Y$ & \#L1                  &                 & \#L2 &                 & \#L3 &                & \#L4 &                 & \#L5 &                \\ \hline
      2   & 2   & 1                     & $\frac12$       & .    &                 & .    &                & .    &                 & .    &                \\
      2   & 3   & 2                     & $\frac12$       & .    &                 & .    &                & .    &                 & .    &                \\
      2   & 4   & 4                     & $\frac12$       & .    &                 & .    &                & .    &                 & .    &                \\
      2   & 5   & 6                     & $\frac12$       & .    &                 & .    &                & .    &                 & .    &                \\
      2   & 6   & 9                     & $\frac12$       & .    &                 & .    &                & .    &                 & .    &                \\
      2   & 7   & 12                    & $\frac12$       & .    &                 & .    &                & .    &                 & .    &                \\
      2   & 8   & 16                    & $\frac12$       & .    &                 & .    &                & .    &                 & .    &                \\
      2   & 9   & 20                    & $\frac12$       & .    &                 & .    &                & .    &                 & .    &                \\
      3   & 3   & 4                     & $\frac12$       & 1    & $\frac{11}{15}$ & .    &                & .    &                 & .    &                \\
      3   & 4   & 11                    & $\frac12$       & 3    & $\frac23$       & .    &                & .    &                 & .    &                \\
      3   & 5   & 18                    & $\frac12$       & 6    & $\frac23$       & .    &                & .    &                 & .    &                \\
      3   & 6   & 29                    & $\frac12$       & 11   & $\frac23$       & .    &                & .    &                 & .    &                \\
      3   & 7   & 42                    & $\frac12$       & 18   & $\frac23$       & .    &                & .    &                 & .    &                \\
      3   & 8   & $\hphantom{^*}$60$^*$ & $\frac12$       & 28   & $\frac23$       & .    &                & .    &                 & .    &                \\
      4   & 3   & 11                    & $\frac12$       & 3    & $\frac23$       & 1    & $\frac56$      & .    &                 & .    &                \\
      4   & 4   & 19                    & $\frac{5}{12}$  & 9    & $\frac58$       & 5    & $\frac34$      & .    &                 & .    &                \\
      4   & 5   & 52                    & $\frac25$       & 28   & $\frac58$       & 13   & $\frac34$      & .    &                 & .    &                \\
      4   & 6   & $\hphantom{^*}$97$^*$ & $\frac25$       & 61   & $\frac58$       & 32   & $\frac34$      & .    &                 & .    &                \\
      5   & 3   & 18                    & $\frac12$       & 6    & $\frac23$       & 2    & $\frac56$      & .    &                 & .    &                \\
      5   & 4   & 52                    & $\frac25$       & 28   & $\frac58$       & 13   & $\frac{8}{11}$ & 2    & $\frac67$       & .    &                \\
      5   & 5   & 72                    & $\frac{15}{38}$ & 51   & $\frac{31}{55}$ & 31   & $\frac{7}{10}$ & 13   & $\frac56$       & .    &                \\
      6   & 3   & 29                    & $\frac12$       & 11   & $\frac23$       & 4    & $\frac56$      & .    &                 & .    &                \\
      6   & 4   & $\hphantom{^*}$97$^*$ & $\frac25$       & 61   & $\frac58$       & 34   & $\frac57$      & 9    & $\frac{13}{16}$ & 2    & $\frac{9}{10}$ \\
      7   & 3   & 42                    & $\frac12$       & 18   & $\frac23$       & 7    & $\frac56$      & .    &                 & .    &                \\
      8   & 3   & $\hphantom{^*}$60$^*$ & $\frac12$       & 28   & $\frac23$       & 12   & $\frac56$      & .    &                 & .    &                \\
      \hline
    \end{tabular}
    }
    \vspace{2pt}
    \subfloat[]{
    \begin{tabular}{|cc|cc|cc|c|}
      \multicolumn{7}{c}{$A=3$ and $B=3$} \\ \hline
      $X$ & $Y$ & \#L1                    &                 & \#L2 &           & $C_\text{max}$ \\ \hline
      2   & 2   & 2                       & $\frac{8}{17}$  & .    &           & 10             \\
      2   & 3   & 22                      & $\frac{8}{17}$  & .    &           & 12             \\
      2   & 4   & 379                     & $\frac{8}{17}$  & .    &           & 16             \\
      3   & 2   & 22                      & $\frac{8}{17}$  & 12   & $\frac34$ & 12             \\
      3   & 3   & $\hphantom{^*}$8747$^*$ & $\frac{13}{31}$ & 8160 & $\frac23$ & 20             \\
      4   & 2   & 379                     & $\frac{8}{17}$  & 337  & $\frac34$ & 16             \\
      \hline
    \end{tabular}
    }
  \end{minipage}%
  \begin{minipage}{0.6\textwidth}
    \centering
    \begin{minipage}{0.5\textwidth}
      \centering
      \subfloat[]{
      \begin{tabular}{|cc|cc|cc|cc|}
        \multicolumn{8}{c}{$A=2$ and $B=3$} \\ \hline
        $X$ & $Y$ & \#L1 &                & \#L2 &                 & \#L3 &           \\ \hline
        2   & 2   & 1    & $\frac12$      & .    &                 & .    &           \\
        2   & 3   & 2    & $\frac12$      & .    &                 & .    &           \\
        2   & 4   & 4    & $\frac12$      & .    &                 & .    &           \\
        2   & 5   & 6    & $\frac12$      & .    &                 & .    &           \\
        2   & 6   & 9    & $\frac12$      & .    &                 & .    &           \\
        2   & 7   & 13   & $\frac12$      & .    &                 & .    &           \\
        3   & 2   & 3    & $\frac12$      & .    &                 & .    &           \\
        3   & 3   & 16   & $\frac{5}{11}$ & 3    & $\frac57$       & .    &           \\
        3   & 4   & 89   & $\frac{7}{16}$ & 26   & $\frac23$       & .    &           \\
        4   & 2   & 13   & $\frac12$      & 4    & $\frac{25}{28}$ & .    &           \\
        4   & 3   & 427  & $\frac{5}{11}$ & 316  & $\frac23$       & 1    & $\frac56$ \\
        5   & 2   & 56   & $\frac12$      & 35   & $\frac{19}{22}$ & .    &           \\
        \hline
      \end{tabular}
      }
      \vspace{2pt}
      \subfloat[]{
      \begin{tabular}{|cc|cc|cc|c|}
        \multicolumn{7}{c}{$A=2$ and $B=4$}  \\ \hline
        $X$ & $Y$ & \#L1                   &                 & \#L2 &                 & $C_\text{max}$ \\ \hline
        2   & 2   & 1                      & $\frac12$       & .    &                 & 5              \\
        2   & 3   & 2                      & $\frac12$       & .    &                 & 5              \\
        2   & 4   & 4                      & $\frac12$       & .    &                 & 5              \\
        2   & 5   & 6                      & $\frac12$       & .    &                 & 5              \\
        3   & 2   & 3                      & $\frac{15}{31}$ & .    &                 & 7              \\
        3   & 3   & 18                     & $\frac37$       & 5    & $\frac{19}{27}$ & 10             \\
        4   & 2   & 16                     & $\frac{15}{31}$ & 4    & $\frac{65}{73}$ & 11             \\
        5   & 2   & $\hphantom{^*}$202$^*$ & $\frac{15}{31}$ & 147  & $\frac{23}{27}$ & 15             \\
        \hline
      \end{tabular}
      }
    \end{minipage}%
    \begin{minipage}{0.5\textwidth}
      \centering
      \subfloat[]{
      \begin{tabular}{|cc|cc|cc|cc|}
        \multicolumn{8}{c}{$A=3$ and $B=2$} \\ \hline
        $X$ & $Y$ & \#L1 &                & \#L2 &           & \#L3 &                 \\ \hline
        2   & 2   & 1    & $\frac12$      & .    &           & .    &                 \\
        2   & 3   & 3    & $\frac12$      & .    &           & .    &                 \\
        2   & 4   & 13   & $\frac12$      & .    &           & .    &                 \\
        2   & 5   & 56   & $\frac12$      & .    &           & .    &                 \\
        3   & 2   & 2    & $\frac12$      & .    &           & .    &                 \\
        3   & 3   & 16   & $\frac{5}{11}$ & 5    & $\frac57$ & .    &                 \\
        3   & 4   & 427  & $\frac{5}{11}$ & 343  & $\frac23$ & .    &                 \\
        4   & 2   & 4    & $\frac12$      & .    &           & .    &                 \\
        4   & 3   & 89   & $\frac{7}{16}$ & 51   & $\frac58$ & 1    & $\frac{13}{16}$ \\
        5   & 2   & 6    & $\frac12$      & .    &           & .    &                 \\
        6   & 2   & 9    & $\frac12$      & .    &           & .    &                 \\
        7   & 2   & 13   & $\frac12$      & .    &           & .    &                 \\
        \hline
      \end{tabular}
      }
      \vspace{2pt}
      \subfloat[]{
      \begin{tabular}{|cc|cc|cc|c|}
        \multicolumn{7}{c}{$A=4$ and $B=2$} \\ \hline
        $X$ & $Y$ & \#L1                   &                 & \#L2 &                 & $C_\text{max}$ \\ \hline
        2   & 2   & 1                      & $\frac12$       & .    &                 & 5              \\
        2   & 3   & 3                      & $\frac{15}{31}$ & .    &                 & 7              \\
        2   & 4   & 16                     & $\frac{15}{31}$ & .    &                 & 11             \\
        2   & 5   & $\hphantom{^*}$202$^*$ & $\frac{15}{31}$ & .    &                 & 15             \\
        3   & 2   & 2                      & $\frac12$       & .    &                 & 5              \\
        3   & 3   & 18                     & $\frac37$       & 7    & $\frac{19}{27}$ & 10             \\
        4   & 2   & 4                      & $\frac12$       & .    &                 & 5              \\
        5   & 2   & 6                      & $\frac12$       & .    &                 & 5              \\
        \hline
      \end{tabular}
      }
    \end{minipage}%
    \vspace{13pt}
    \begin{minipage}{\textwidth}
      \centering
      \subfloat[]{
      \begin{tabular}{|cccc|cc|cc|c|}
        \multicolumn{9}{c}{Remaining scenarios} \\ \hline
        $X$ & $Y$ & $A$ & $B$ & \#L1 &                 & \#L2 &                 & $C_\text{max}$ \\ \hline
        2   & 2   & 3   & 4   & 2    & $\frac{5}{11}$  & .    &                 & 10             \\
        2   & 2   & 4   & 4   & 3    & $\frac37$       & .    &                 & 17             \\
        2   & 2   & 5   & 5   & 4    & $\frac{16}{41}$ & .    &                 & 26             \\
        2   & 3   & 3   & 4   & 28   & $\frac{5}{11}$  & .    &                 & 14             \\
        2   & 3   & 4   & 3   & 104  & $\frac{5}{11}$  & .    &                 & 19             \\
        3   & 2   & 3   & 4   & 104  & $\frac{5}{11}$  & 86   & $\frac{11}{15}$ & 19             \\
        3   & 2   & 4   & 3   & 28   & $\frac{5}{11}$  & 18   & $\frac{11}{15}$ & 14             \\
        \hline
      \end{tabular}
      }
    \end{minipage}%
  \end{minipage}%
  \caption{
  Number of classes of ENS boxes which violate different communication scenarios.
  For each scenario, we provide a list of classes; completeness is proved in most cases and conjectured in the cases marked with a star.
  Next to the number of boxes that cannot be reproduced with communication $d$, we indicate the minimum visibility with respect to the corresponding polytope, that is, the ratio of the LHV+$d$it value to the NS one.
  We provide $C_\text{max}$, the largest clique size we found in the scenario.
  For $A\cdot B\leq6$, we have $C_\text{max}=5$.
  For concision, we only provide results for one version of each set of two party-symmetric scenarios.
  }
  \label{tab:comm}
\end{table*}

Now we prove the assertion in the main text concerning the arbitrary communication complexity of NS strategies.
In the scenario $(m,m,2,2)$, recall that
\begin{equation}
  p=\left(\begin{array}{c|c|c|c}
    C      & C      & \cdots & C      \\ \hline
    C      & A      & \cdots & C      \\ \hline
    \vdots & \vdots & \ddots & \vdots \\ \hline
    C      & C      & \cdots & A      \\
  \end{array}\right),
\end{equation}
where
\begin{equation}
  C=\begin{pmatrix}
    \frac12 & 0       \\
    0       & \frac12
  \end{pmatrix}
  \quad\text{and}\quad
  A=\begin{pmatrix}
    0       & \frac12 \\
    \frac12 & 0
  \end{pmatrix},
\end{equation}
and define $F=4p-1$.
In other words, $F$ has entries
\begin{equation}\nonumber
  F(ab|xy)=\left\{\begin{array}{cl}
    1  & \text{if }a+b=(1-\delta_{x,1})\delta_{x,y}\pmod2, \\
    -1 & \text{otherwise.}
  \end{array}\right.
\end{equation}

For $d\leq m$, we seek the LHV+$d$it bound for the inequality $F$.
Without loss of generality, we can restrict to the case where Alice's deterministic strategy is to always return her first output.

If $d=m$, then it is trivial that the NS value $m^2$ can be achieved, as Alice can simply communicate her input to Bob, who then has complete information and can score one for the inequality $F$ for all values of $x$ and $y$.

If $d<m$, then, by symmetry of $F$, the choice of the communication function is only relevant with respect to the first input, and it is easy to see that one optimal strategy is to communicate 1 for $x\in [1,m-d+1]$ and $x-d$ for $x\in [m-d+2,m]$.
When Bob receives 1, he is indeed left with a reduced version of the inequality $F$ (of size $m-d+1$), for which his optimal strategy is to always return his first output, only losing when $x=y\in [2,m-d+1]$; otherwise, he has perfect knowledge.
Combining both paths, we obtain the stated value of $m^2-2(m-d)$.

\end{document}